\documentclass[twocolumn,superscriptaddress]{revtex4-2}

\RequirePackage{graphicx}

\usepackage{amsmath}
\usepackage{amssymb}
\usepackage{booktabs}
\usepackage{float}
\usepackage{gensymb}
\usepackage{graphicx}
\usepackage{hyperref}
\usepackage{placeins}
\usepackage{soul}
\usepackage{upgreek}
\usepackage{xcolor}
\usepackage{xfrac}

\begin{document}

\title{Novel Approach for Evaluating Detector-Related Uncertainties in a LArTPC Using MicroBooNE Data}
% List of institutions in command form:
\newcommand{\Bern}{Universit{\"a}t Bern, Bern CH-3012, Switzerland}
\newcommand{\BNL}{Brookhaven National Laboratory (BNL), Upton, NY, 11973, USA}
\newcommand{\UCSB}{University of California, Santa Barbara, CA, 93106, USA}
\newcommand{\Cambridge}{University of Cambridge, Cambridge CB3 0HE, United Kingdom}
\newcommand{\CIEMAT}{Centro de Investigaciones Energ\'{e}ticas, Medioambientales y Tecnol\'{o}gicas (CIEMAT), Madrid E-28040, Spain}
\newcommand{\Chicago}{University of Chicago, Chicago, IL, 60637, USA}
\newcommand{\Cincinnati}{University of Cincinnati, Cincinnati, OH, 45221, USA}
\newcommand{\CSU}{Colorado State University, Fort Collins, CO, 80523, USA}
\newcommand{\Columbia}{Columbia University, New York, NY, 10027, USA}
\newcommand{\Edinburgh}{University of Edinburgh, Edinburgh EH9 3FD, United Kingdom}
\newcommand{\FNAL}{Fermi National Accelerator Laboratory (FNAL), Batavia, IL 60510, USA}
\newcommand{\Granada}{Universidad de Granada, Granada E-18071, Spain}
\newcommand{\Harvard}{Harvard University, Cambridge, MA 02138, USA}
\newcommand{\IIT}{Illinois Institute of Technology (IIT), Chicago, IL 60616, USA}
\newcommand{\KSU}{Kansas State University (KSU), Manhattan, KS, 66506, USA}
\newcommand{\Lancaster}{Lancaster University, Lancaster LA1 4YW, United Kingdom}
\newcommand{\LANL}{Los Alamos National Laboratory (LANL), Los Alamos, NM, 87545, USA}
\newcommand{\Manchester}{The University of Manchester, Manchester M13 9PL, United Kingdom}
\newcommand{\MIT}{Massachusetts Institute of Technology (MIT), Cambridge, MA, 02139, USA}
\newcommand{\Michigan}{University of Michigan, Ann Arbor, MI, 48109, USA}
\newcommand{\Minnesota}{University of Minnesota, Minneapolis, MN, 55455, USA}
\newcommand{\NMSU}{New Mexico State University (NMSU), Las Cruces, NM, 88003, USA}
\newcommand{\Oxford}{University of Oxford, Oxford OX1 3RH, United Kingdom}
\newcommand{\Pitt}{University of Pittsburgh, Pittsburgh, PA, 15260, USA}
\newcommand{\Rutgers}{Rutgers University, Piscataway, NJ, 08854, USA}
\newcommand{\SLAC}{SLAC National Accelerator Laboratory, Menlo Park, CA, 94025, USA}
\newcommand{\SDSMT}{South Dakota School of Mines and Technology (SDSMT), Rapid City, SD, 57701, USA}
\newcommand{\Maine}{University of Southern Maine, Portland, ME, 04104, USA}
\newcommand{\Syracuse}{Syracuse University, Syracuse, NY, 13244, USA}
\newcommand{\TelAviv}{Tel Aviv University, Tel Aviv, Israel, 69978}
\newcommand{\Tennessee}{University of Tennessee, Knoxville, TN, 37996, USA}
\newcommand{\UTA}{University of Texas, Arlington, TX, 76019, USA}
\newcommand{\Tufts}{Tufts University, Medford, MA, 02155, USA}
\newcommand{\VTech}{Center for Neutrino Physics, Virginia Tech, Blacksburg, VA, 24061, USA}
\newcommand{\Warwick}{University of Warwick, Coventry CV4 7AL, United Kingdom}
\newcommand{\Yale}{Wright Laboratory, Department of Physics, Yale University, New Haven, CT, 06520, USA}
%%\newcommand{\listerThanks}{Now at University of Wisconsin, Madison}

% So that institutions appear in alphabetical order:
\affiliation{\Bern}
\affiliation{\BNL}
\affiliation{\UCSB}
\affiliation{\Cambridge}
\affiliation{\CIEMAT}
\affiliation{\Chicago}
\affiliation{\Cincinnati}
\affiliation{\CSU}
\affiliation{\Columbia}
\affiliation{\Edinburgh}
\affiliation{\FNAL}
\affiliation{\Granada}
\affiliation{\Harvard}
\affiliation{\IIT}
\affiliation{\KSU}
\affiliation{\Lancaster}
\affiliation{\LANL}
\affiliation{\Manchester}
\affiliation{\MIT}
\affiliation{\Michigan}
\affiliation{\Minnesota}
\affiliation{\NMSU}
\affiliation{\Oxford}
\affiliation{\Pitt}
\affiliation{\Rutgers}
\affiliation{\SLAC}
\affiliation{\SDSMT}
\affiliation{\Maine}
\affiliation{\Syracuse}
\affiliation{\TelAviv}
\affiliation{\Tennessee}
\affiliation{\UTA}
\affiliation{\Tufts}
\affiliation{\VTech}
\affiliation{\Warwick}
\affiliation{\Yale}

% Authors in alphabetical order
%\author{P.~Abratenko} \affiliation{\Michigan} % for CC incl and any papers using MCS
\author{P.~Abratenko} \affiliation{\Tufts} 
\author{R.~An} \affiliation{\IIT}
\author{J.~Anthony} \affiliation{\Cambridge}
\author{L.~Arellano} \affiliation{\Manchester}
\author{J.~Asaadi} \affiliation{\UTA}
\author{A.~Ashkenazi}\affiliation{\TelAviv}
\author{S.~Balasubramanian}\affiliation{\FNAL}
\author{B.~Baller} \affiliation{\FNAL}
\author{C.~Barnes} \affiliation{\Michigan}
\author{G.~Barr} \affiliation{\Oxford}
% ...

\collaboration{The MicroBooNE Collaboration}
\thanks{microboone\_info@fnal.gov}\noaffiliation
% if the above does not work for your version of LATeX, please try the version below
%\email[]{microboone\_info@fnal.gov}
\date{\today}

\begin{abstract}
Primary challenges for current and future precision neutrino experiments using liquid argon time projection chambers (LArTPCs) include understanding detector effects and quantifying the associated systematic uncertainties. This paper presents a novel technique for assessing and propagating LArTPC detector-related systematic uncertainties. The technique makes modifications to simulation waveforms based on a parameterization of observed differences in ionization signals from the TPC between data and simulation, while remaining insensitive to the details of the detector model. The modifications are then used to quantify the systematic differences in low- and high-level reconstructed quantities. This approach could be applied to future LArTPC detectors, such as those used in SBN and DUNE.
\end{abstract}

\maketitle

\section{Introduction}
\label{sec:intro}

In a modern particle physics experiment, simulation of the detector response is used to estimate efficiencies and resolutions of measured quantities.
These efficiencies and resolutions are necessary in order to fully interpret the data produced by the experiment.
The possible differences between what is simulated and the actual detector response therefore lead to bias on physics measurements. This potential bias is quantified in the form of detector systematic uncertainties.
This paper describes a method in which the response of the MicroBooNE LArTPC detector~\cite{detector} is characterized in data and simulation. The results are used to modify simulated signals to thereby produce samples of modified simulated events.
Comparisons between modified simulations and the nominal simulation can be used to identify measurement biases and to estimate detector systematic uncertainties.
Understanding detector effects and systematic uncertainties is critical for achieving the physics goals of future LArTPC-based experiments, such as SBN~\cite{sbn} and DUNE~\cite{dune}. The detector-related uncertainties must be reduced to the level of a few percent and estimated precisely to reach the design sensitivities.

The principal detector of MicroBooNE is a wire-based liquid argon time projection chamber (TPC) with a single drift region.
The trajectories of charged particles through the liquid argon are detected by drifting ionization electrons in an electric field to three parallel planes of sense wires.
The drifted ionization charge measured at the wire planes is sensitive to a number of known detector effects, such as electron--ion recombination~\cite{icarusrecomb,recomb}, electron diffusion~\cite{bnldiffusion,icarusdiffusion,diffusion}, space charge effects~\cite{laser,sce_cosmic}, and electron attenuation~\cite{elifetime,calib}. It is also subject to effects related to the model that describes the induced signal on the wires due to the drifting electrons and the electronics response~\cite{sp1,sp2}.
These effects are difficult to disentangle.

The method detailed in this paper is used to address systematic uncertainties related to ionization charge in the TPC that can be described by changes in the amplitude and width of signals on the wires.
This method produces a set of simulations where the signals on wires are modified---differences between these varied simulation sets and the nominal simulation are taken as an estimate of the uncertainty on the nominal simulation's modeling of the detector response to ionization.
For the subset of the detector variations where this approach can be used, it has two significant advantages over modeling-based estimates.
First, by working with digitized wire waveforms in both data and simulation, this procedure does not depend explicitly on the modeling used for different components of detector response simulation.
It therefore captures residual effects that are not well-described by existing detector models or that are not fully simulated, providing a more robust, data-driven assessment of uncertainties related to the detector model.
Second, it is relatively computationally efficient.
By directly modifying waveforms, this approach avoids the computationally intensive steps of simulating the drifting of ionization electrons and deconvolving the resulting signals.
As a result, the method outlined is about an order of magnitude faster than running the full simulation each time.

The structure of the paper is as follows: Section~\ref{sec:overview} gives a brief overview of the method, including a description of the relevant detector variables and the parameters that are used to characterize the detector's response.
Section~\ref{sec:samples} defines the event samples in data and simulation.
Section~\ref{sec:ratios} describes the procedure for extracting the data-to-simulation comparisons, which take the form of ratios of waveform properties.
Section~\ref{sec:wiremod} describes the application of these ratios to modifying the wire waveforms.
Section~\ref{sec:impacts} presents the results of applying this method to higher-level reconstructed quantities.
Section~\ref{sec:future} discusses the potential improvements and extensions.
Section~\ref{sec:summary} presents the summary and conclusion.

\section{Overview of Method}
\label{sec:overview}

The MicroBooNE detector is a liquid argon time projection chamber (LArTPC) designed to observe neutrino interactions. It is located on-axis along the Booster Neutrino Beam (BNB)~\cite{bnb} at Fermilab, and is also exposed to an off-axis flux of the Neutrinos from the Main Injector (NuMI) beam~\cite{numi}. Compared to the BNB beam, the NuMI beam is higher in energy and has a larger electron neutrino contribution.

\begin{figure}[ht]
    \centering
    \includegraphics[width=\columnwidth]{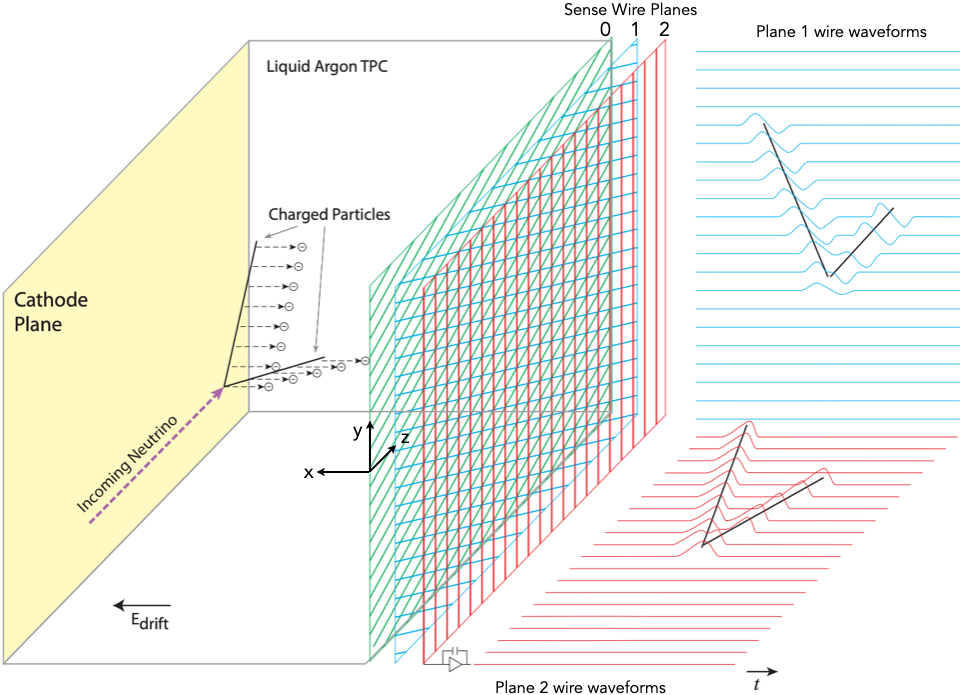}
    \caption{The MicroBooNE detector and operating principles, adapted from Ref.~\cite{detector}, as described in the text. The green and blue wire planes are the induction planes; the red wire plane is the collection plane. The right-hand portion of the figure shows the wire waveforms before deconvolution.}
    \label{fig:detector}
\end{figure}

When charged particles traverse the detector, they deposit energy that liberates ionization electrons and also produces prompt vacuum ultraviolet scintillation photons.
The ionization electrons drift in the applied electric field until they reach the three sense wire planes located at the anode, as illustrated in Figure~\ref{fig:detector}.
The electrostatic potentials of the wire planes are set up such that ionization electrons pass through the first two wire planes before ultimately ending their trajectory on a wire in the last plane. The drifting electrons induce signals on the first two planes, referred to as induction planes (planes 0~and~1), and additively constitute the signals in the final plane, referred to as the collection plane or plane~2. The collection plane wires are aligned vertically and the induction plane wires are oriented at $\pm 60\degree$ from the vertical.
The voltage of each wire is digitized by on-detector electronics, and recorded over time  to produce raw waveforms.
To process recorded raw waveforms offline, a noise-filtering algorithm is applied~\cite{noise} and then the field responses are removed from the signals via a Gaussian deconvolution process~\cite{sp1,sp2} to produce a waveform that measures the charge that arrived at each wire as a function of time.
Scintillation photons are observed by an array of 32~photo-multiplier tubes (PMTs) located behind the wire planes. The optical information is used for triggering the detector.

The detector's response to an ionizing particle depends on the position and the amount of energy deposited, as well as the angular orientation of the particle's trajectory with respect to the wires~\cite{sp1,sp2}. 
The MicroBooNE coordinate system is defined such that the $x$ axis points along the drift electric field direction from the anode to the cathode, the $y$ axis points vertically up, and the $z$ axis points along the BNB beam direction to complete a right-handed coordinate system.
As the response depends on the orientation of a particle trajectory, it is useful to define the detector angles $\theta_{XZ}$ and $\theta_{YZ}$ for a displacement vector $\Updelta \vec{r_i}$ with components $(\Updelta x_i, \Updelta y_i, \Updelta z_i)$ as below.
\begin{equation}
    \begin{aligned}
        \theta_{XZ,i} &= \arctan(\Updelta x_i / \Updelta z_i) \\
        \theta_{YZ,i} &= \arctan(\Updelta y_i / \Updelta z_i)
    \end{aligned}
    \label{eqn:def_thetas}
\end{equation}
In the coordinate system used, the direction of the BNB has both $\theta_{XZ}$ and $\theta_{YZ}$ equal to zero.
The vector, $\Updelta \vec{r_i}$, is taken to be the (true) local direction of travel of the simulated particle that produced a particular wire waveform.

Later, in Section~\ref{sec:ratios}, ``rotated'' angles relevant for the two induction planes are introduced.
The detector response is characterized as a function of these five variables: $x$, $y$, $z$, $\theta_{XZ}$, and $\theta_{YZ}$.
Much of the variability in the detector's response in $y$ and $z$ is driven by the presence of non-responsive wires in one plane, which can affect the behavior of the signals on nearby wires on the other planes~\cite{sp2}.
The different planes have different orientations in the $yz$-plane, but the locations of wire-crossings are at fixed points this 2D plane; for this reason $y$ and $z$ are considered together.
The remaining variables are considered independently.

The effects of each of the variables on the post-deconvolution wire waveforms are described in terms of a Gaussian fit to the waveform, called a \textit{hit}. A hit has an integrated charge $Q$, proportional to the number of ionization electrons that produced the wire signal, and a width $\sigma$, measured in waveform time ticks. A tick corresponds to 0.5~$\mu \rm{s}$ as defined by the 2~MHz sampling rate of the ADCs~\cite{detector}.
To quantify how the wire waveforms differ between data and simulation, the differences are expressed as data-to-simulation ratios.

The hits are used as the basis to apply the modifications to the underlying waveforms.
Digitized waveforms from each wire in each event are divided into wire signal regions separated by signal-free regions, which are zero-suppressed.
Each wire signal region can be described by the sum of one or more Gaussian functions with some peak position, integrated charge, and width.
Each constituent Gaussian function is modified according to the properties of the simulated energy deposits that are matched to it, by applying the data-to-simulation differences provided by the ratio functions for $Q$ and $\sigma$ for each detector variable.
The technical details are described in Sec.~\ref{sec:wiremod}.

The variation as a function of $x$ position captures the dependence of the signal width on, for example, the charge cloud spreading out (diffusion), and of the signal amplitude on electrons being absorbed by impurities (attenuation).
The local variation in $y$ and $z$ can account for the distortion of the signal due to deviations in the electric field between the wire planes resulting from non-responsive and cross-connected wires.
The variations in the angular variables $\theta_{XZ}$ and $\theta_{YZ}$ can describe distortions in the waveforms due to imperfect modeling of the signals that drift charge induces on the wires and of the electronics response. This is particularly relevant for extended charge distributions, because the response can include interference between signals induced by different parts of the charge distribution on the same wire. This interference depends on the angular orientation of the charge distribution relative to the wire planes in a way that is challenging to model precisely~\cite{sp1,sp2}.
All of these waveform-level modifications are agnostic to the downstream reconstruction and analysis chain as well as the upstream detector simulation model.
For evaluating the full range of systematic uncertainties related to the MicroBooNE detector, separate variations are considered for the drift electric field model~\cite{laser,sce_cosmic}, the electron--ion recombination model parameters~\cite{calib}, and the scintillation light model parameters.

\section{Data and Simulation Event Samples}
\label{sec:samples}

To determine the hit properties (integrated charge and width), cosmic ray muon tracks are used. They provide an abundant and well-understood event sample in which each of the five relevant position and angular variables can be reconstructed.
The data tracks are selected from beam-off data, which is collected using the same optical trigger as the beam-on data but when there is no neutrino beam (so-called ``beam-off'' events).
The triggered beam-off data comes from MicroBooNE's Run 1 period, taken between February and October 2016.
It was verified that consistent results were obtained using different run periods, so for simplicity the measurements are made using Run 1 and applied to all other runs.

The simulation tracks are selected from a sample of single muons that are generated using CORSIKA~\cite{corsika}.
The signals from these simulated muons are overlaid on cosmic data that is collected using a random (unbiased) trigger when there is no neutrino beam.
The cosmic data overlay incorporates the detector noise and cosmic muon backgrounds found in data events. 
This technique is also applied to the simulated neutrino events discussed in Sec.~\ref{sec:impacts}.
The unbiased cosmic data used in this procedure comes from the run period that matches the data sample to which the simulation is being compared.
For the simulated muon samples used to measure the hit properties, this means unbiased beam-off data from the Run 1 period is used.

\begin{figure} 
    \centering
    \includegraphics[width=\columnwidth]{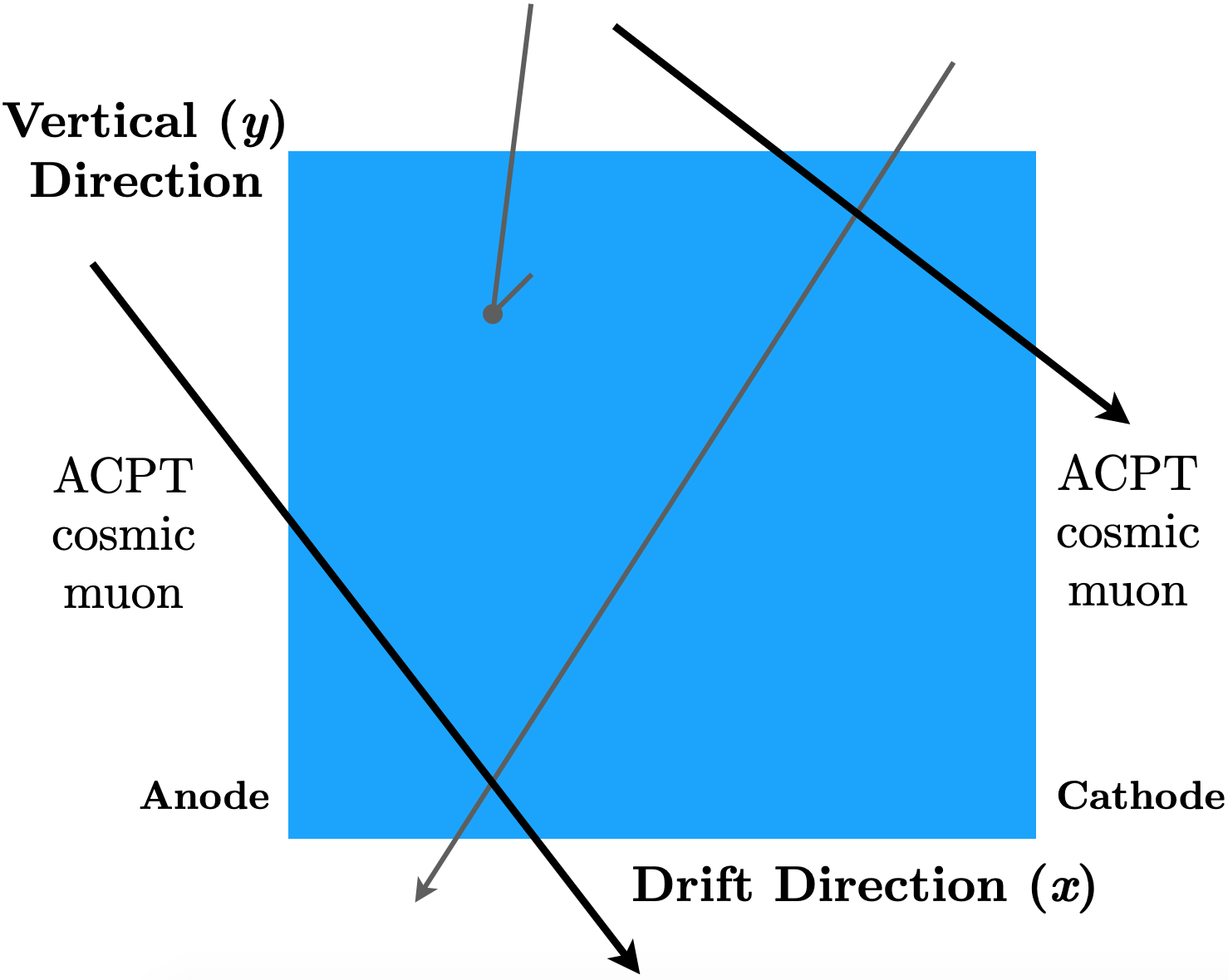}
    \caption{Illustration of two examples of anode/cathode piercing tracks (ACPT), shown in black. The track must cross at least one of the anode or the cathode. The other tracks, shown in gray, are cosmic muons that do not satisfy the ACPT criteria.}
    \label{fig:acpt}
\end{figure}

The $x$ position of an energy deposit in the MicroBooNE TPC is determined from the drift time of its ionization tracks relative to the trigger time of the event combined with measurements of the local drift velocity~\cite{laser,sce_cosmic}.
To reconstruct the $x$ position of a given particle track, it is therefore necessary to match that track to a flash of scintillation light, whose offset from the trigger time is readily known.
This is achieved by using cosmic tracks that are topologically consistent with having crossed the anode or the cathode in-time with the flash of scintillation light that triggered the beam-off event. In addition, the opposite end of the track is required to have crossed either the opposite face of the detector or the top or bottom. These are called anode/cathode piercing tracks (ACPT) and are illustrated in Figure~\ref{fig:acpt}.

\subsection{Reconstruction}
\label{subsec:reconstruction}

The Pandora multi-algorithm package~\cite{pandora} is used to reconstruct 3D tracks from the ionization charge collected at the wires. 
These tracks are then matched to the flash of scintillation light, collected by the PMT system, which triggered the TPC readout~\cite{ccnp}.
If the track has an ACPT topology and matched to the scintillation light which triggered the detector, it is selected. These types of tracks have little ambiguity in the TPC-to-PMT matching, leading to a sample that is very pure in tracks with the correct $x$ position assigned.
Selected tracks are corrected for spatial distortions due to nonuniform electric fields in the detector~\cite{laser,sce_cosmic}.

Based on simulation studies, more than 95\% of the selected track candidates are true ACPT tracks with correctly determined $x$ positions. Additionally, such through-going cosmic muon tracks generally behave as minimally ionizing particles along their entire length and therefore make a good ``standard candle'' of ionization per unit track length.
Note that the geometrical requirements of this selection combined with the fact that cosmic muons are mostly downward-going mean that ACPT muon trajectories tend to populate the regions near the anode (low $x$ position) and cathode (high $x$ position).

\section{Measuring the Detector Response}
\label{sec:ratios}

Using the cosmic ray muon ACPT samples described above in Section~\ref{sec:samples}, the method proceeds by determining the dependence of the two hit properties on each of the five geometric variables stated.
With a measurement of these dependencies made in both data and simulation in each variable, the ratio of the two is formed and used as a measure of the scale of the discrepancy between them. This section details the determination of these ratios. The ratios will later be used (see Sec.~\ref{sec:wiremod}) to derive modifications to the wire waveforms that capture differences due to detector modeling.

\subsection{Measurements in $x$}
\label{subsec:x_ratio}

First, consider variations in charge response as a function of the $x$ position. This is sensitive to drift-dependent effects, such as electron diffusion and attenuation.
To measure the response, all hits associated with reconstructed ACPT muon tracks are used to form distributions of the hit charge and hit width across bins in $x$ position.
The detector is divided into bins in $x$ using a variable binning scheme to ensure a reasonable number of entries in each of the $x$ bins. ACPT trajectories are concentrated near the anode and the cathode, so the bins are narrower in those regions.
The binning is determined separately for each of the wire planes.
Each bin contains hits from several thousand ACPT muons.

\begin{figure*}[ht]
    \centering
    \includegraphics[width=0.95\columnwidth]{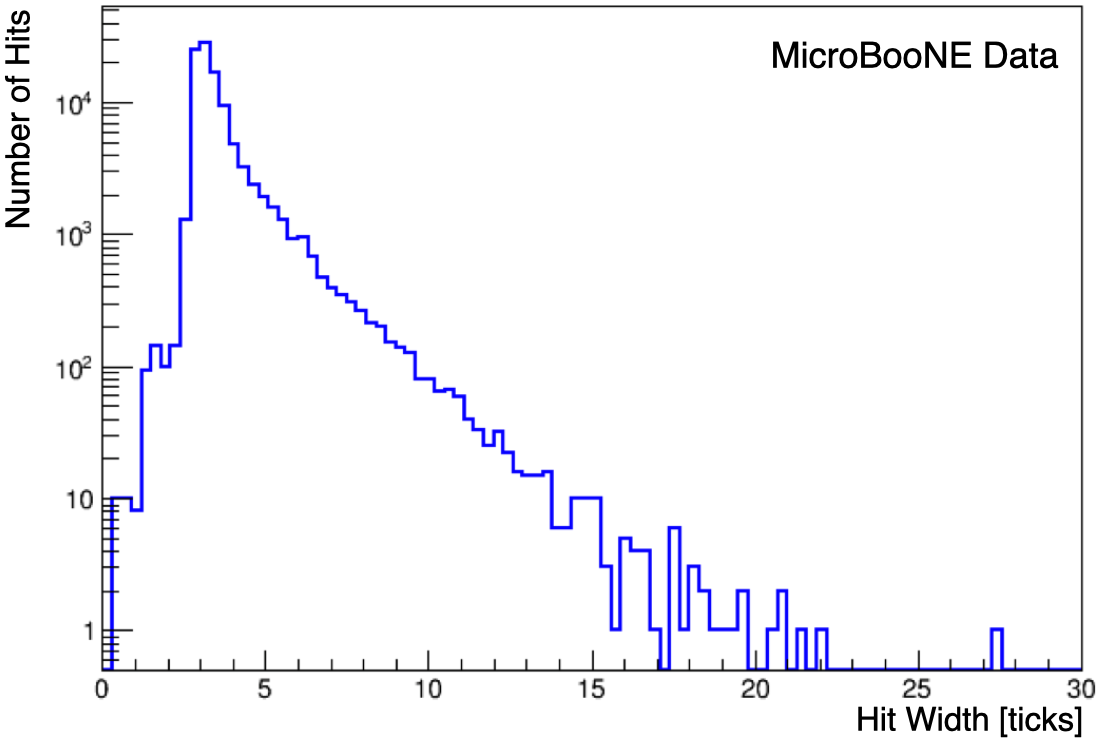}
    \caption{Distribution of hit widths on the collection plane for $1.6 < x < 4.3~\text{cm}$ in the cosmic data. The spread in the distribution is driven by other sources of variability, such as the position in the $yz$-plane and the angular orientation of the track. The distribution is asymmetric and is not well described by any simple analytic function. This motivates the specialized algorithm based on the iterative truncated mean that is described in the text. A tick corresponds to 0.5~$\mu \rm{s}$ of time~\cite{detector}.}
    \label{fig:prop_in_x_bin}
\end{figure*}

\begin{figure*}[ht]
    \centering
    \includegraphics[width=0.95\columnwidth]{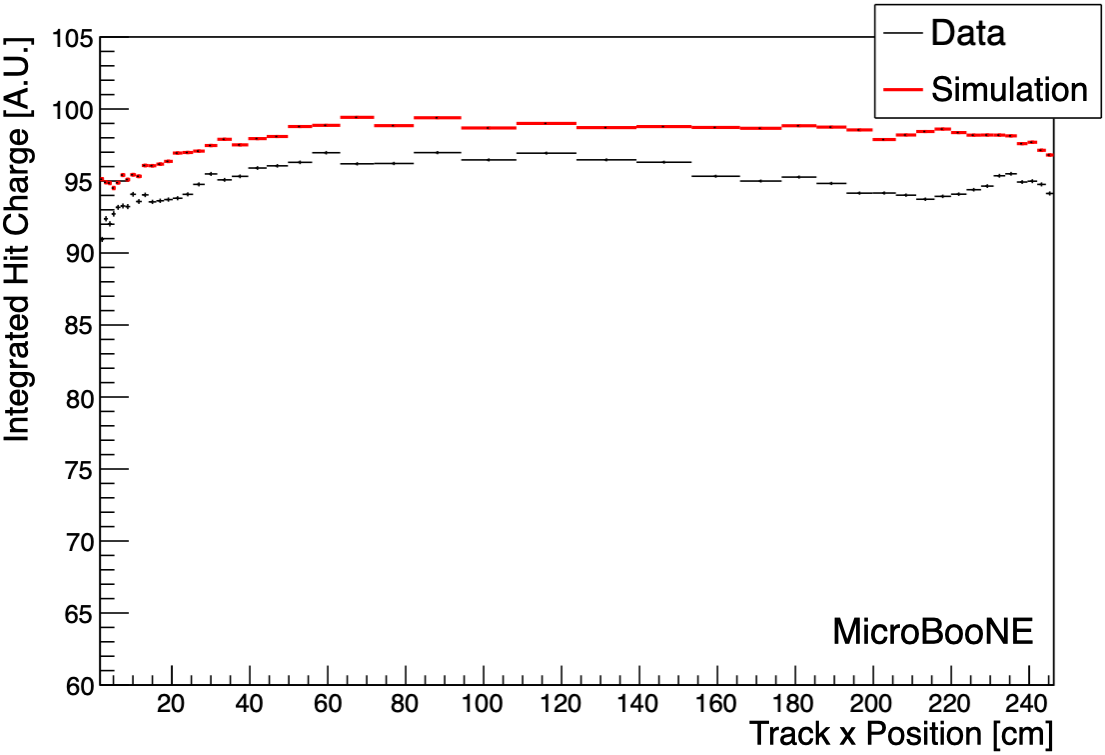}
    $\qquad$
    \includegraphics[width=0.95\columnwidth]{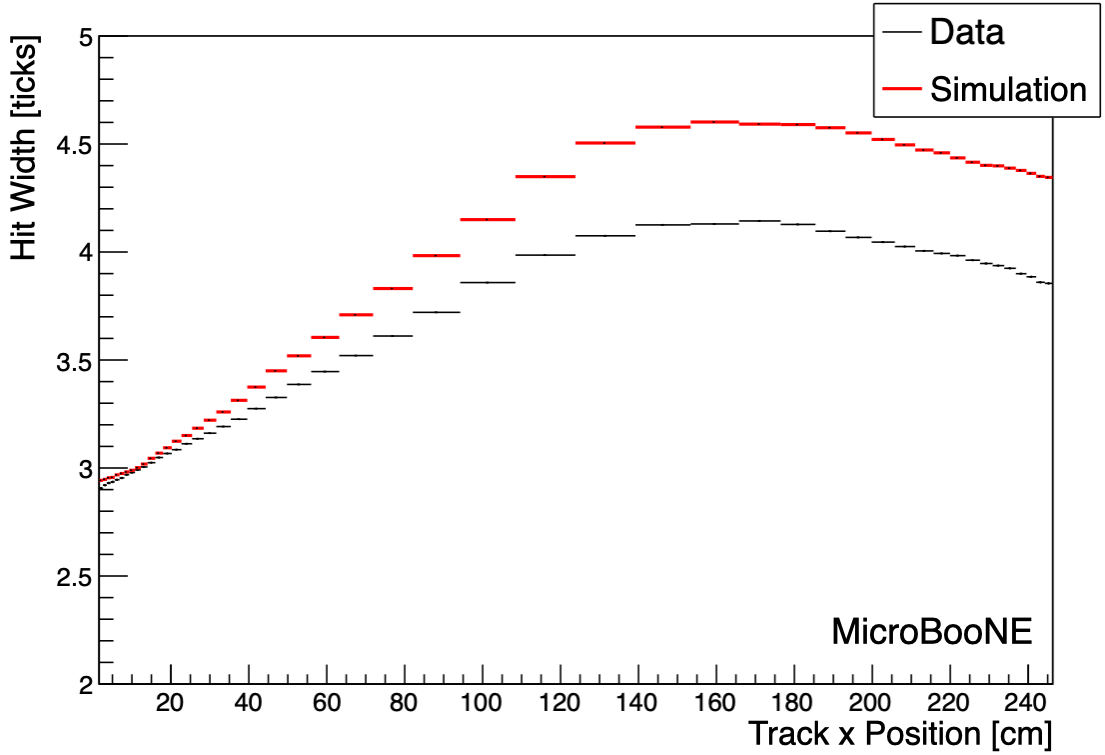}
    \caption{Hit charge and hit width vs.\ $x$ in data and simulation (MC) for the collection plane. The values are computed from histograms similar to the example shown in Figure~\ref{fig:prop_in_x_bin} using the algorithm based on the iterative truncated mean described in the text.}
    \label{fig:prop_vs_x_prof}
\end{figure*}

Within each bin, the values of the hit properties have some intrinsic spread due to the different positions and orientations of tracks, as demonstrated in the distribution of hit widths of a typical $x$ bin in Figure~\ref{fig:prop_in_x_bin}.
To facilitate the measurement of the variation that is due to the $x$ position, the peaks of the hit charge and width distributions in each $x$ bin are calculated using an iterative truncated mean algorithm.
The algorithm starts with all the hits in the bin and computes the mean, the median, and the standard deviation. Hits that are more than 2 standard deviations below the median or more than 1.75 standard deviations above it are removed, and all quantities are then recalculated. The boundaries for the truncation reflect the asymmetry of the underlying distributions, and were empirically determined to improve the accuracy and stability of the peak finding algorithm. This step is repeated until the calculated mean meets the convergence criteria of changing by less than $10^{-4}$. The resulting distribution for means of hit charge and width from the collection wire plane are shown in Figure~\ref{fig:prop_vs_x_prof}.

\begin{figure*}[ht]
    \centering
    \includegraphics[width=0.95\columnwidth]{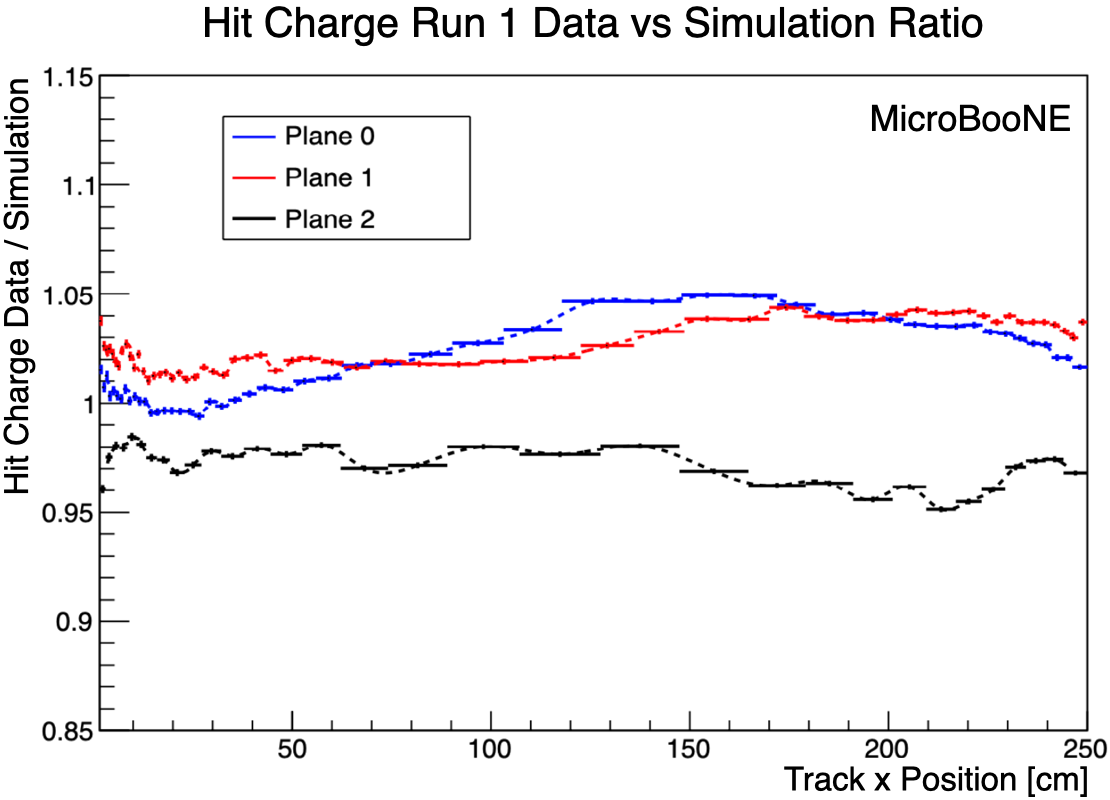}
    $\qquad$
    \includegraphics[width=0.95\columnwidth]{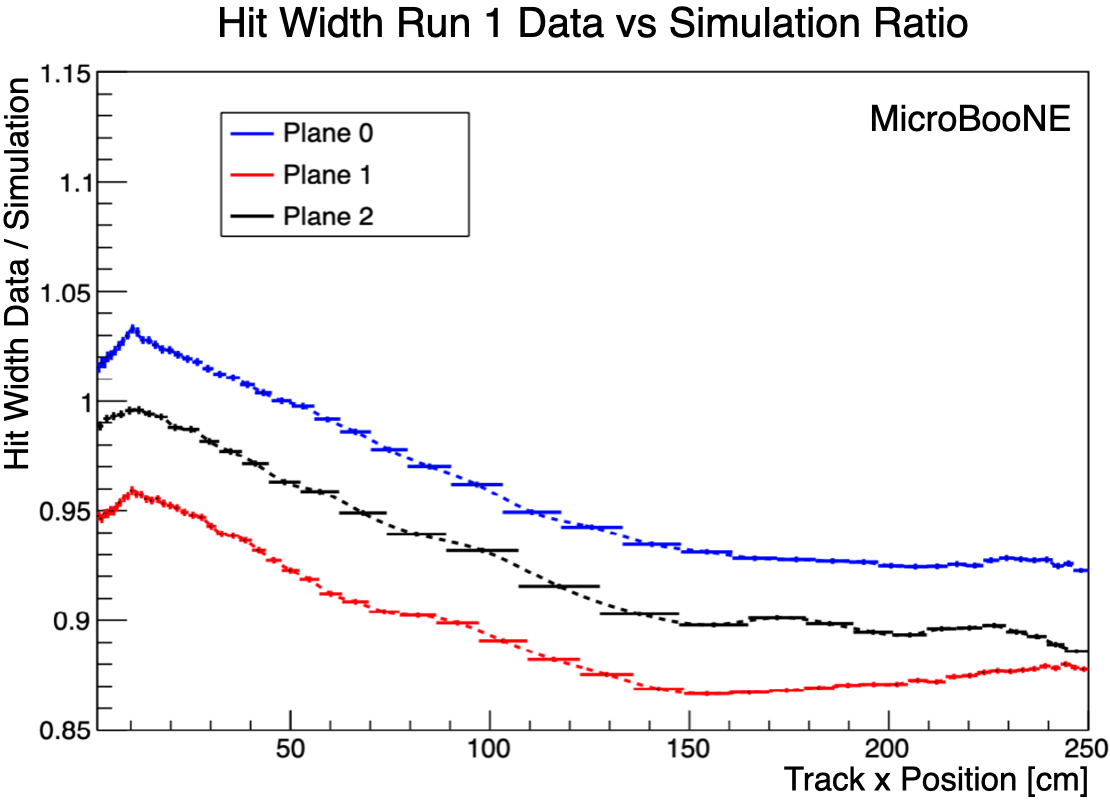}
    \caption{Ratios (data/simulation) and fitted simulation modification functions for mean hit charge and mean hit width vs.\ $x$ on each of the three wire planes. The solid lines are the bin values, with error bars showing the statistical uncertainties, and the dashed lines are spline fits. The width of each bin is indicated by the solid horizontal bars. The binning is chosen to ensure high statistics in each bin.}
    \label{fig:prop_vs_x_ratio}
\end{figure*}

The ratio of the typical hit properties in data to simulation is computed in each bin in $x$ using the peaks found by the truncated mean algorithm.
A spline fit to the measured ratio is performed to obtain a smooth function that describes the data-to-simulation differences, as shown in Figure~\ref{fig:prop_vs_x_ratio}.
This fit provides the simulation modification factor.

\begin{figure*}[ht]
    \centering
    \includegraphics[height=1.85in]{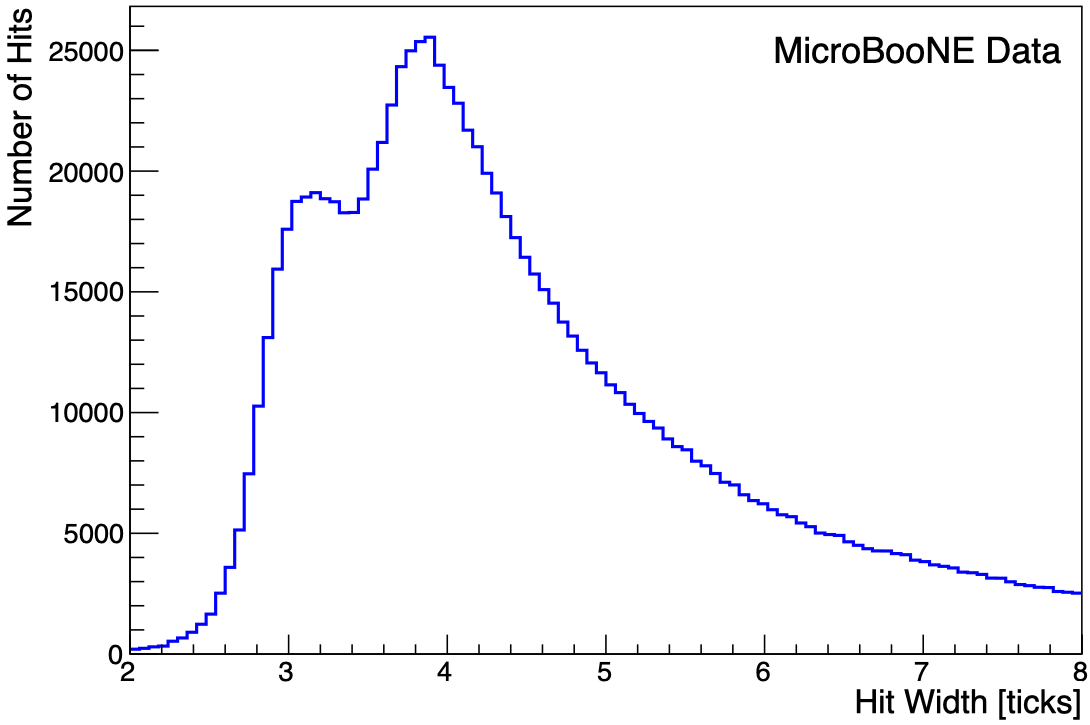}
    \includegraphics[height=1.85in]{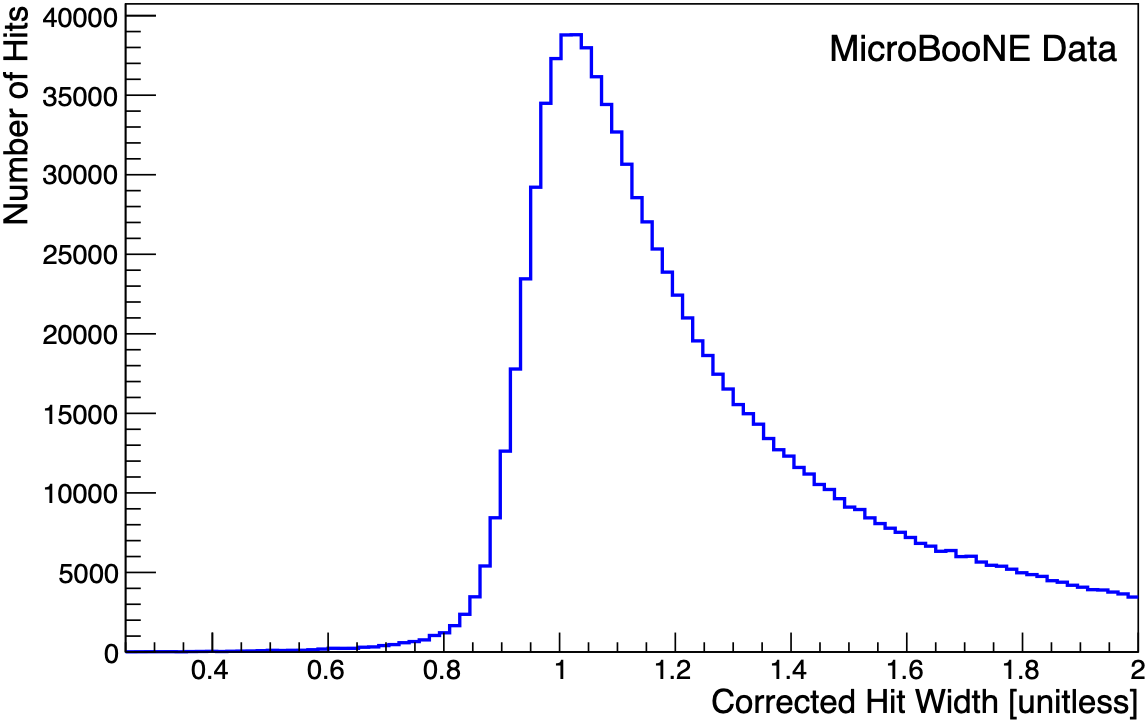}
    \caption{Distribution of hit widths on the collection plane for $-10 < y < 0$\,cm in the cosmic data. On the left, this distribution before any correction for the hit width dependence on $x$. A time tick translates to 0.5 $\mu \rm{s}$~\cite{detector}. The ``double-peak'' structure is evident, where the low-width peak comes from ACPT trajectory points near the anode and the high-width peak comes from points near the cathode (see Figure~\ref{fig:prop_vs_x_prof}). On the right, the $x$-correction has been applied and the double-peak structure is removed. In this case, the hit widths (in ticks) have been divided by the median hit width at the corresponding $x$ position (in ticks), so the resulting quantity is dimensionless.}
    \label{fig:double_banding}
\end{figure*}

\subsection{An $x$ Correction for Other Measurements}
\label{subsec:x_corr}

The hit widths (and to a lesser extent the charges) have large variations as a function of $x$, specifically between the cathode and anode. As shown in Figure~\ref{fig:prop_vs_x_prof}, the measured hit widths vary by up to 50\% across the drift direction. As a result, the hit widths have broad distributions when projected onto the other four geometric variables.
For the ACPT muon sample in particular, where the trajectories tend to populate the regions at high and low $x$, this leads to a ``double-peak'' structure in the hit width distributions in both data and simulation. This complicates the measurement of the hit width dependence as a function of these other variables, as the truncated mean is no longer a well-behaved estimate of the peak position.
An example of this double-peak feature for a bin in $y$ is shown in Figure~\ref{fig:double_banding}.
To account for this, the measurements for the other variables are based on hit properties that have been corrected for their known $x$-dependence.

Spline fits to the results in Figure~\ref{fig:prop_vs_x_prof}, for data and simulation and for each wire plane separately, provide expected hit properties for a hit at a given $x$ position, on a given plane, in data or simulation.
Each hit's charge and width is then divided by the relevant expected value to produce ``$x$-corrected'' hit properties.
This process produces distributions of corrected hit properties that have a median value of one, by construction.

The remaining measurements in $(y,z)$ and the angular angular variables use these $x$-corrected hit properties.
As well as avoiding the difficulties with the double-peak structure, this process removes any global offsets from the remaining measurements, placing all global scalings in the $x$-dependence.
The remaining measurements are shape-only in their respective variables.
These measurements are further described in the sections below.

\subsection{Measurements in $(y, z)$}
\label{subsec:yz_ratio}

Next consider the behavior of hit charge and width in the $yz$-plane.
The detector effects that dominate the behavior in these two variables are TPC channels that are shorted or cross-connected, which distorts the electric field between the wire planes and therefore the wire response~\cite{sp2}. This creates local nonuniformities in the charge response in $(y,z)$.
Note that the detector response in the nominal simulation incorporates a data-driven tuning for this effect.
This section will briefly describe the method for tuning the simulation, followed by the method for extracting the residual difference that will be used to evaluate an uncertainty.

First, the nominal simulation is tuned by scaling the simulated local $(y, z)$ charge response based on measurements of the charge deposited per unit track length, $dQ/dx$.
The median $dQ/dx$ is measured in $5 \times 5$~cm$^2$ bins over the $yz$-plane. This is used to calculate the following quantity in each $(y,z)$ bin for each wire plane in data and simulation:
\begin{equation}
    C_{(y_i,z_i)} = \frac{(dQ/dx)_{\text{global}}}{(dQ/dx)_{(y_i,z_i)}},
\end{equation}
where $(dQ/dx)_\text{global}$ is the global median $dQ/dx$ value of the entire $(y,z)$ plane and $(dQ/dx)_{(y_i,z_i)}$ is the local median in $(y,z)$ bin $i$.
The simulated charge response is scaled by the ratio of $C_{(y_i,z_i)}$ measured in data to the one measured in simulation for each wire plane.
The reconstructed $dQ/dx$ quantities are generally corrected for these local nonuniforimities using the $C_{(y_i,z_i)}$ values from data as part of the downstream analysis~\cite{calib}.
However, the reconstructed quantities used for the technique described in this paper are Gaussian fits to the deconvolved waveforms, where the $yz$-plane uniformity calibration is not applied.

The method described in this paper is used to measure the residual bias in the model for the nonuniformities in the tuned simulation.
The same sample of ACPT muons and the peak-finding algorithm as described in Section~\ref{subsec:x_ratio} are employed, but with the $x$-correction described in Section~\ref{subsec:x_corr} applied to the hit properties.
The $(y,z)$ bins are optimized in 2D to again ensure reasonable numbers of entries in each. The result is a set of rectangular $(y,z)$ bins that vary in size based on the density of hits on each wire plane (typically about 4--5~$\text{cm}$ on each side) and contain hits from at least a thousand ACPT muons.

\begin{figure*}
    \centering
    \includegraphics[width=0.4\textwidth]{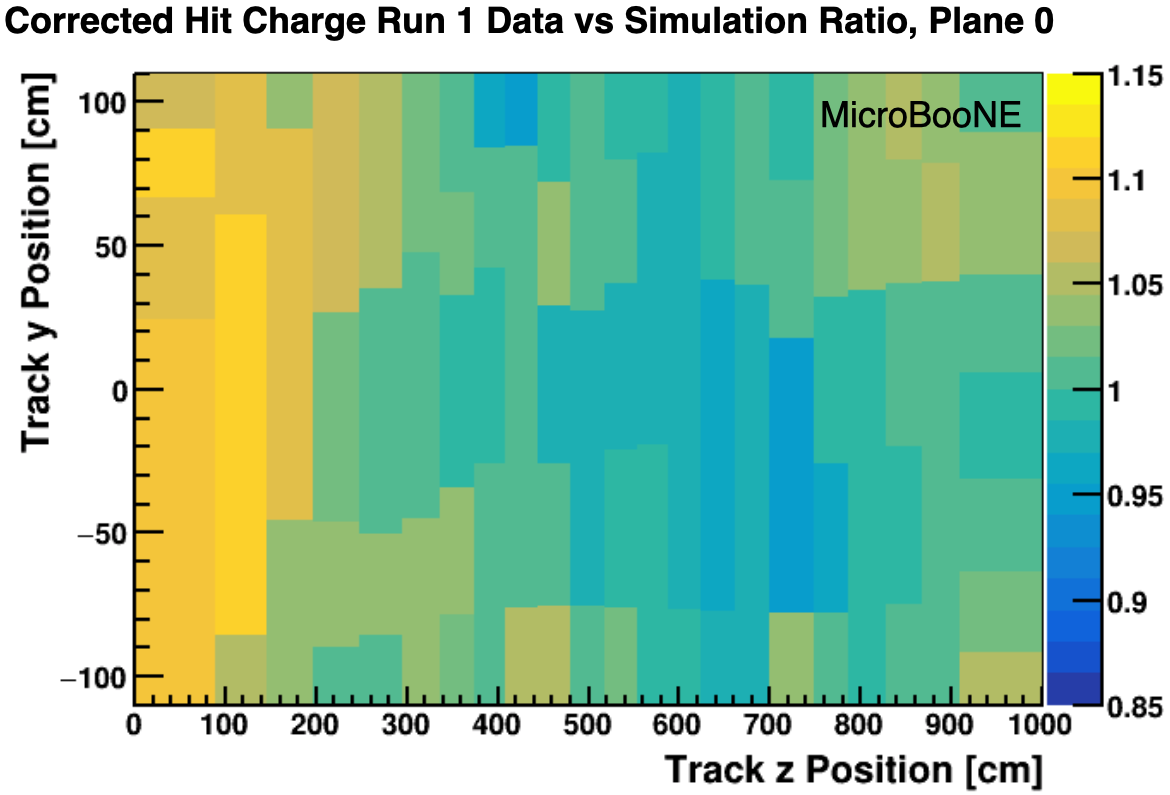}
    \includegraphics[width=0.4\textwidth]{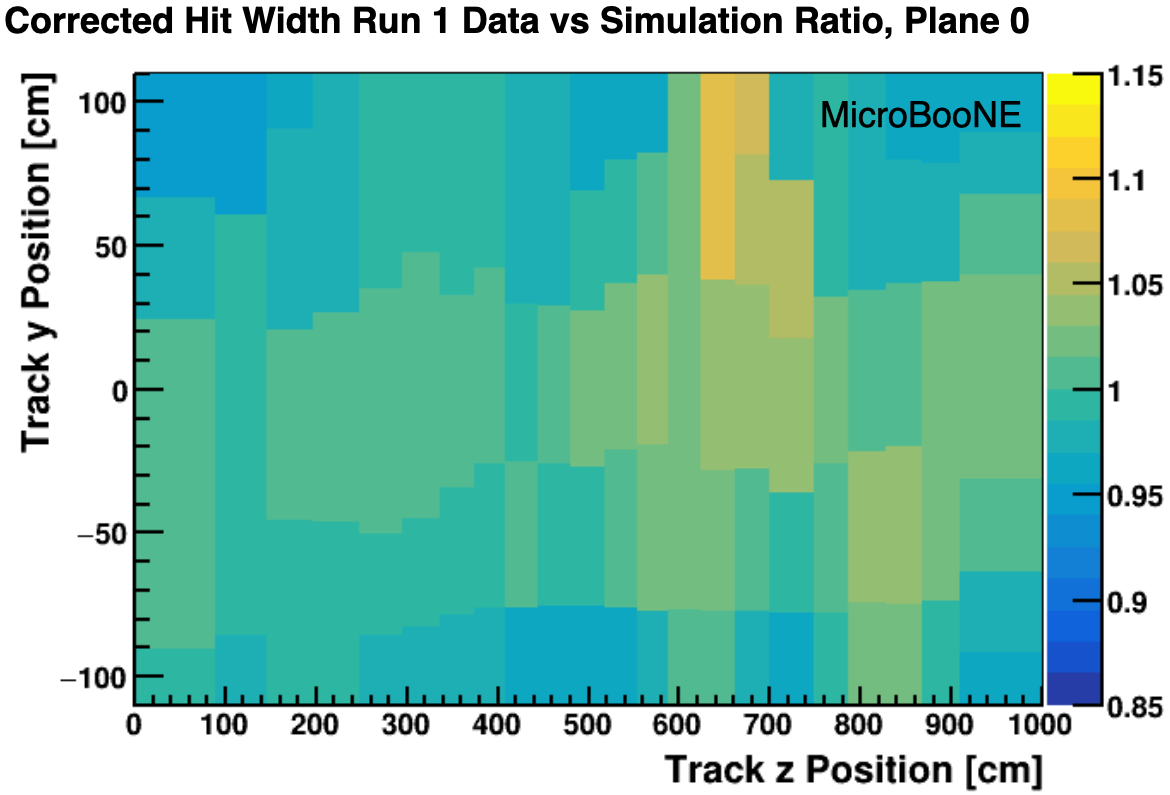}
    \includegraphics[width=0.4\textwidth]{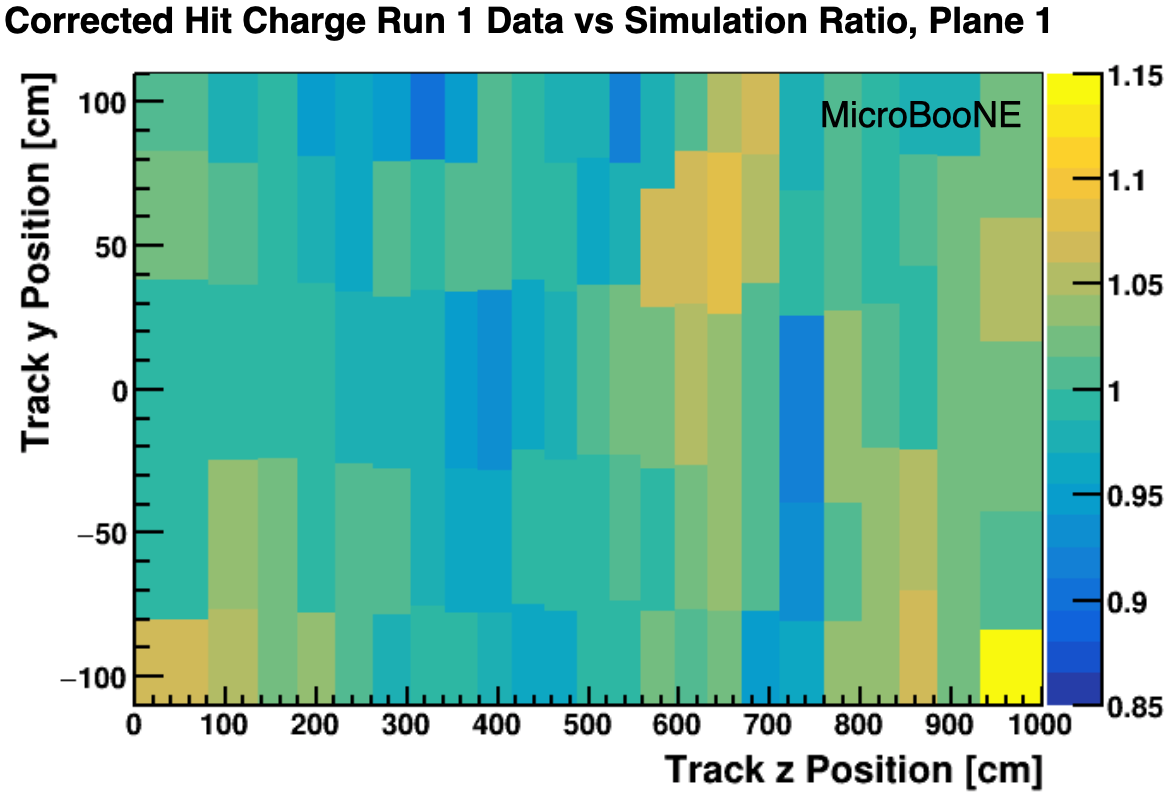}
    \includegraphics[width=0.4\textwidth]{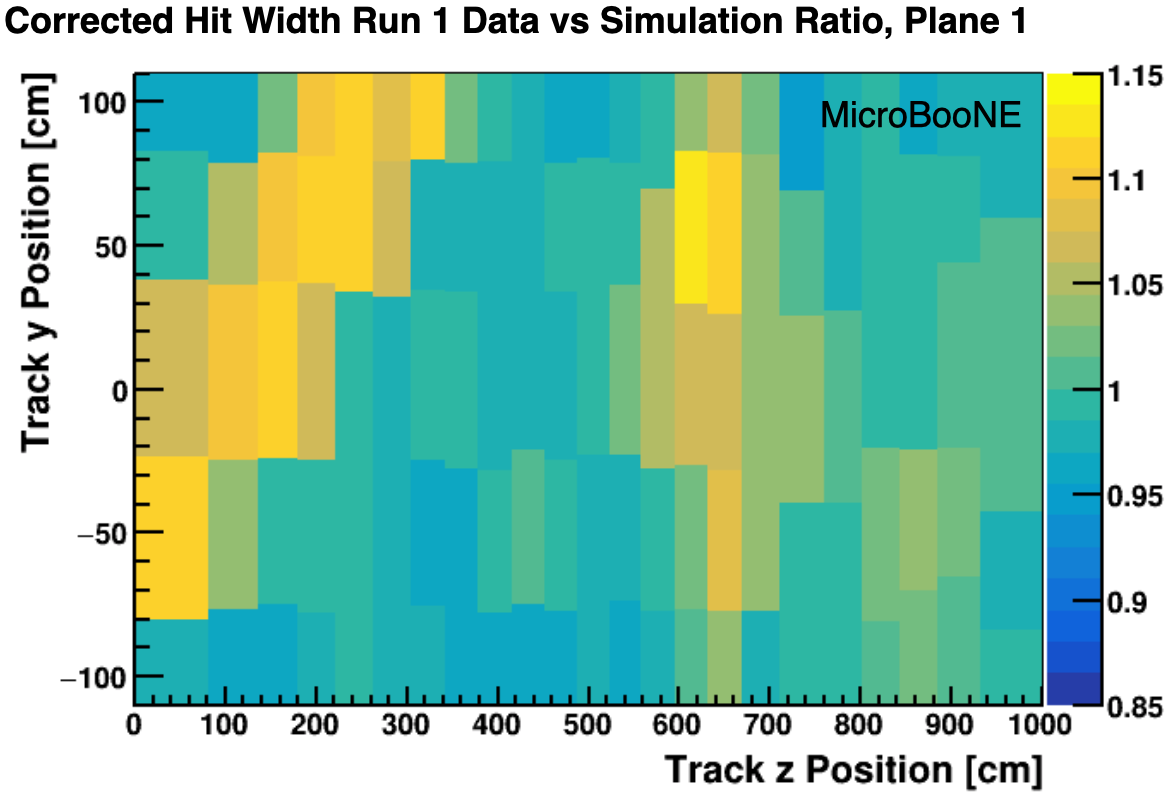}
    \includegraphics[width=0.4\textwidth]{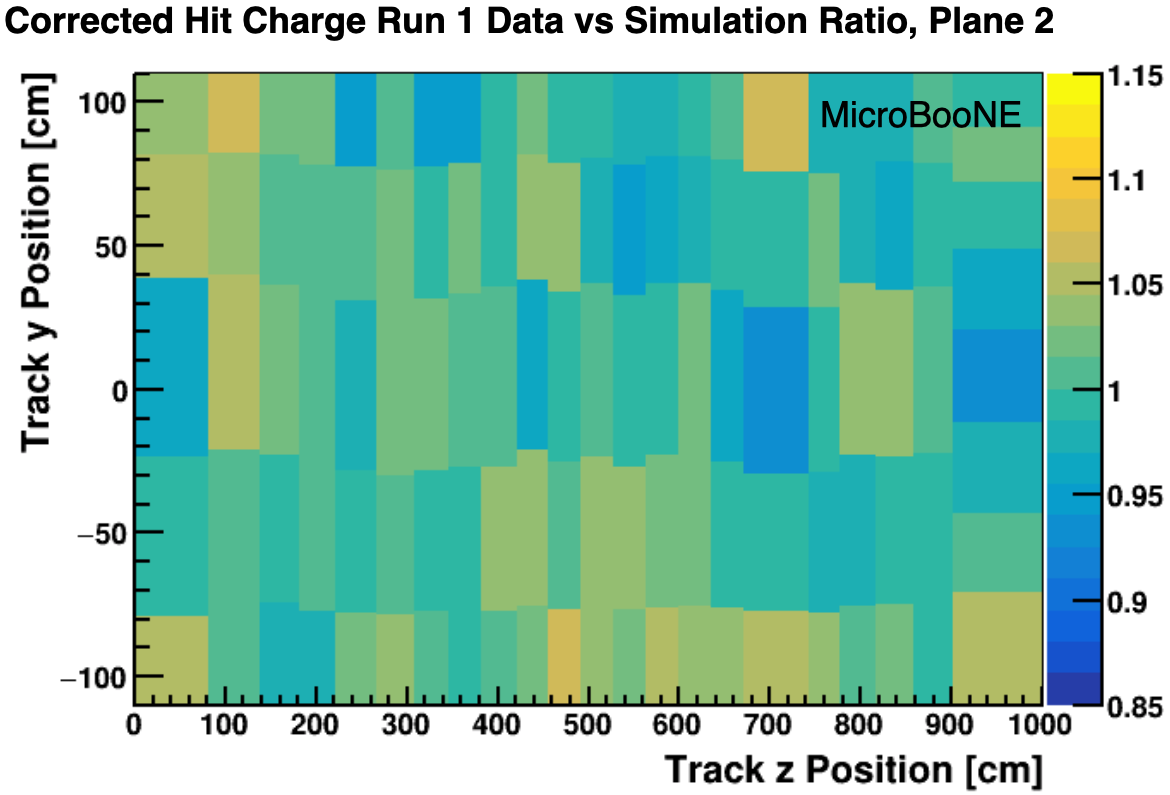}
    \includegraphics[width=0.4\textwidth]{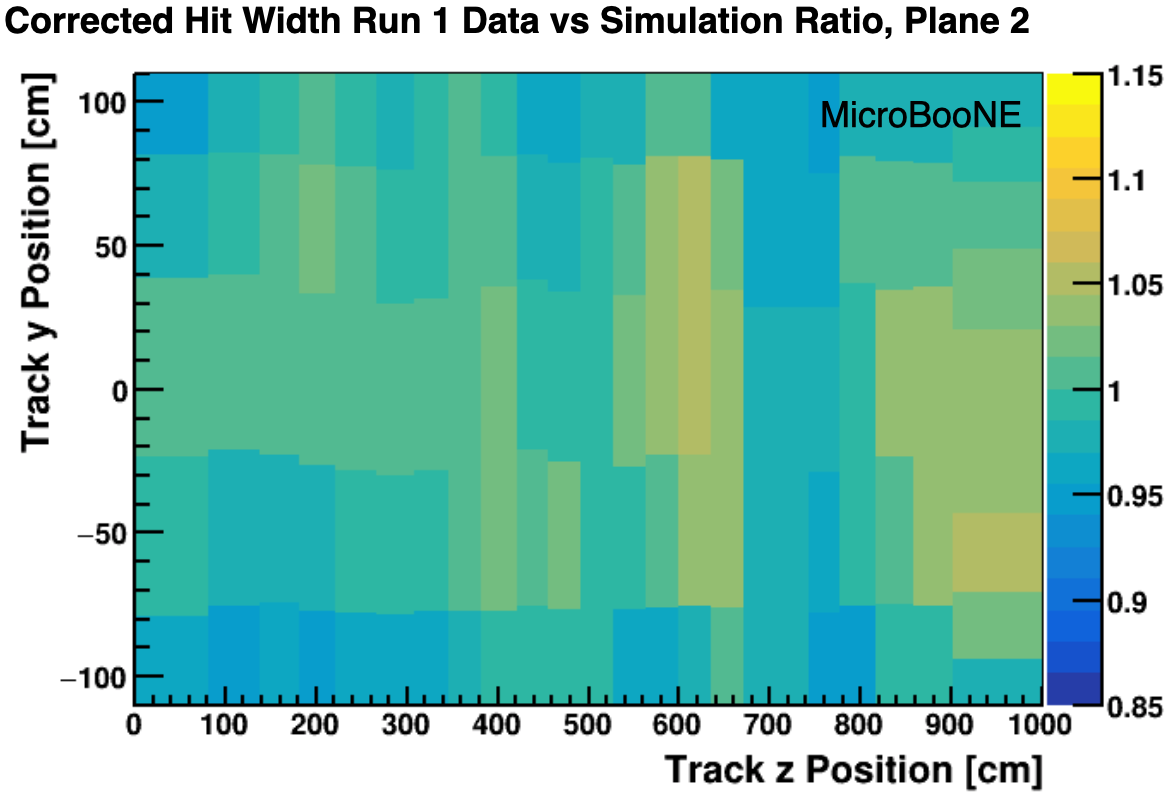}
    \caption{Ratios (data/simulation) for hit charge and width vs.\ $(y,z)$. The left column shows the hit charge; the right column shows the hit widths. The top row shows the ratios on the first induction plane; the middle row shows the ratios on the second induction plane; and the bottom row shows the ratios on the collection plane. Note the color axis is the same on all six graphs.}
    \label{fig:prop_vs_yz_ratio}
\end{figure*}

Figure~\ref{fig:prop_vs_yz_ratio} shows the results of applying the procedure outlined above.
A smooth function of $y$ and $z$ that describes these ratios is obtained by interpolating between points in the 2D space. In the interior of the detector, the points are the centers of the $(y,z)$ bins. For bins where one edge is along the boundary of the detector, an additional point is placed at the midpoint of that edge with the same value as the point at the center of the bin. Additional points are placed in the four corners of the $(y, z)$ plane, with values given by the ratio at the center of the corner bin.

\subsection{Measurements in Angular Variables}
\label{subsec:angle_ratios}

In addition to the position of the charge in the detector discussed in the preceding sections, this method also considers the orientation of the particle trajectory in angular variables. This captures effects related to long-range induced charge signals on the wires as well as the signal processing.
The same procedure as in the previous section is applied, including the $x$-correction for the hit properties described in in Section~\ref{subsec:x_corr}.
This section details some special considerations related to the choice of basis for the angular variables, and how to handle angles relative to each wire plane where signal processing and hit finding become less reliable.

The two angles most relevant for describing the detector response to a charged particle track are the angle with respect to the drift direction ($x$) and the angle with respect to the wire direction (which is different for each wire plane).
For the collection plane, where the wires are oriented vertically, these are the angles $\theta_{XZ}$ and $\theta_{YZ}$, respectively, as defined in Equation~\ref{eqn:def_thetas}.
For the induction planes, where the wires are oriented at $\pm 60\degree$ from the vertical, analogous angles are defined with respect to a different set of basis vectors, $x'$, $y'$, and $z'$, where $x'$ remains the drift direction, $y'$ is the appropriate wire direction, and $z'$ completes an orthogonal right-handed basis.
Mathematically, this is expressed by the following expressions for the first (upper sign) and second (lower sign) induction planes.
\begin{equation}
\label{eqn:def_rotate}
  \begin{aligned}
      x' &= x \\
      y' &= y\cos(60\degree) \pm z\sin(60\degree) \\
      z' &= y\sin(60\degree) \mp z\cos(60\degree) \\
    \end{aligned}
\end{equation}
The angles $\theta_{XZ}$ and $\theta_{YZ}$ are used for all wire planes with the understanding that these quantities always refer to the angle definition relevant for the plane in question.
With this choice of angular basis, the variations in hit properties in $\theta_{XZ}$ and $\theta_{YZ}$ can be treated independently.

It was verified that the detector response in both integrated charge and width does not depend on the quadrant for these angles, i.e.\ that the wire response is independent of the particle's direction (up vs.\ down, etc.), as expected.
Because of this, it is possible to ``fold'' all angles into the space between 0 and $\pi/2$.

Using this angular basis, the variations in the $x$-corrected properties of the hits are measured as a function of angles.
The ACPT muons do not have an isotropic angular distribution, so a variable binning scheme is employed here.
The peak in each angular bin in data and simulation is computed using the same algorithm described in Section~\ref{subsec:x_ratio}.
However, as either $\theta_{XZ}$ or $\theta_{YZ}$ approach $\pi/2$, the corresponding deconvolved waveform is no longer well-described by a single Gaussian function, and is instead an extended charge deposition~\cite{sp1}.
Above 1.4~radians (about $80\degree$), the observed distribution of hit charges and widths cannot be reliably used to characterize the detector's response.
The simulation modification factor in this bin ($R_N$) is instead extrapolated using the maximum absolute difference from $1.0$ over the rest of the angular space ($\Updelta R_\text{max}$) while maintaining the sign of the difference from the adjacent measured bin ($R_{N-1}$):
\begin{equation}
    \begin{aligned}
    \Updelta R_\text{max} &= \max_{\text{bins $k$}} \left| R_k - 1 \right| \\
    R_N &= 1 + \left( \text{sign}(R_{N-1}-1) \cdot \Updelta R_\text{max} \right).
    \end{aligned}
    \label{eqn:extrapolate}
\end{equation}

It is worth noting that a displacement vector with $\theta_{XZ}=\pi/2$ or $\theta_{YZ}=\pi/2$ also has zero $z$-component. In the MicroBooNE coordinate system, the BNB points along the $z$ direction, so the region in which we use this extrapolation is perpendicular to the neutrino beam.

\begin{figure}[h]
    \centering
    \includegraphics[width=0.95\columnwidth]{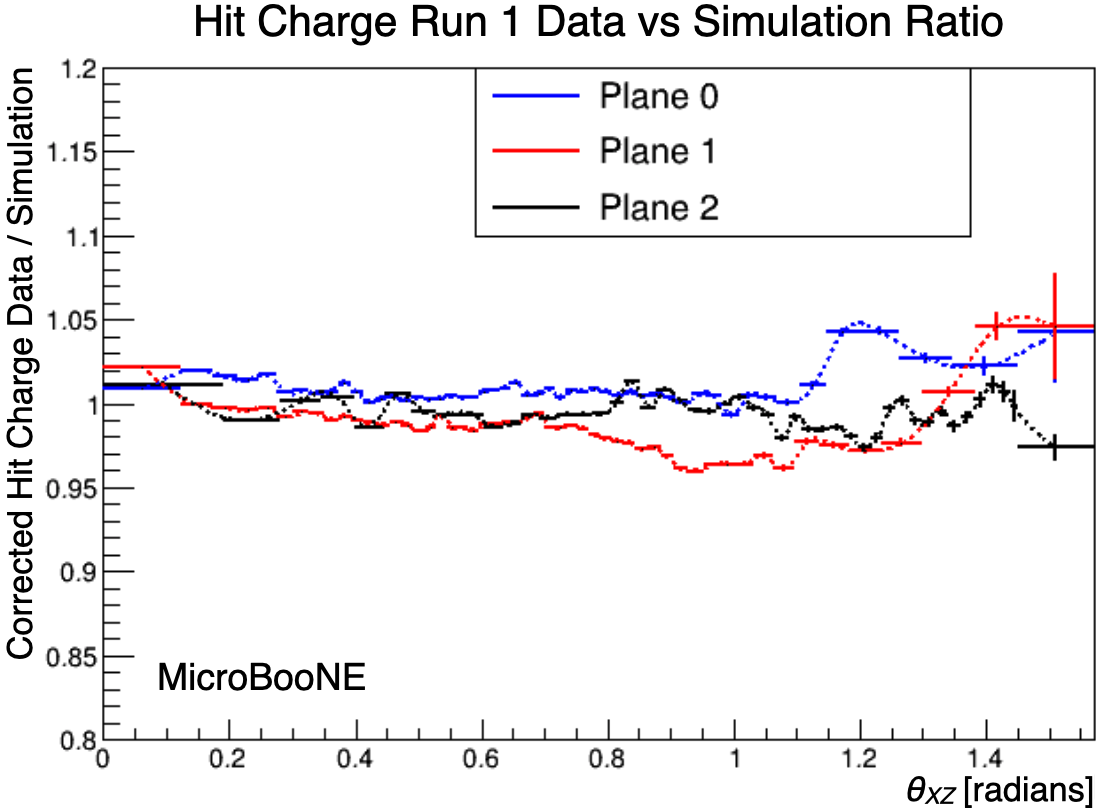}
    \includegraphics[width=0.95\columnwidth]{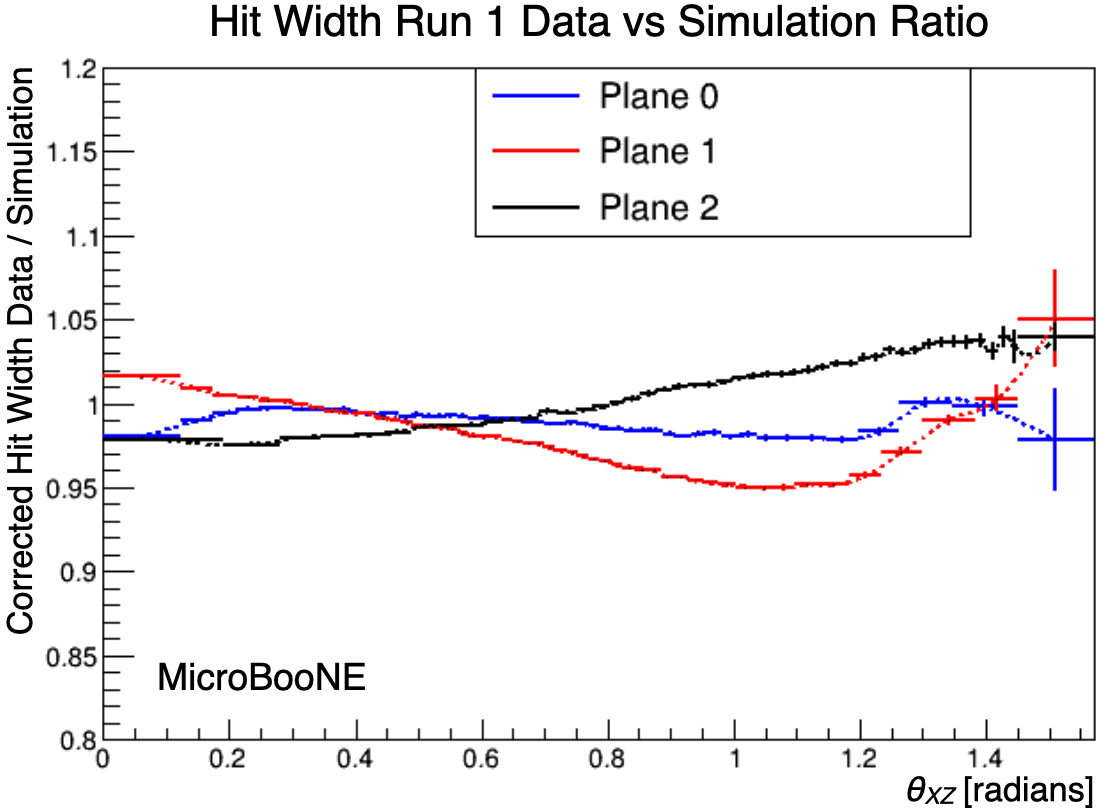}
    \caption{Ratio functions (data/simulation) for hit charge (top) and hit width (bottom) vs.\ $\theta_{XZ}$. The solid lines are the values of the ratio in each bin, and the dashed lines are the spline fits. The bins at $1.4 < \theta_{XZ} < \pi/2~\text{rad}$ are extrapolated as described in the text.}
    \label{fig:prop_vs_thetaXZ_ratio}
\end{figure}

\begin{figure}[h]
    \centering
    \includegraphics[width=0.95\columnwidth]{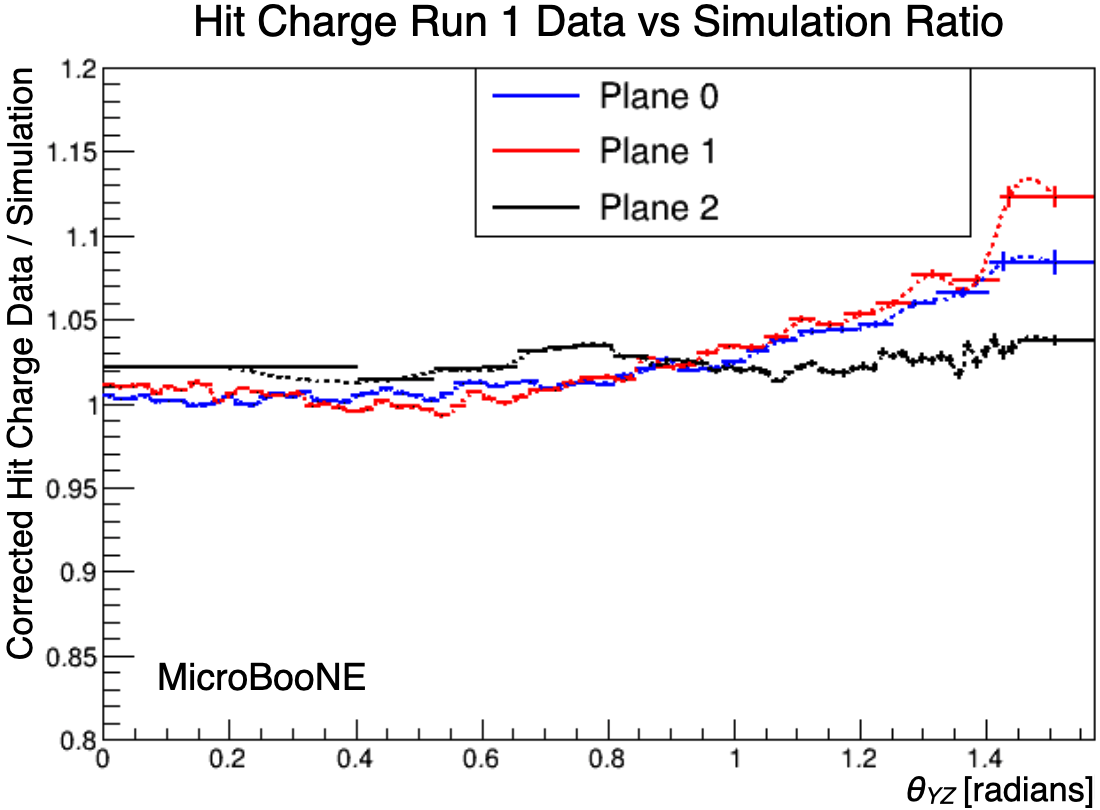}
    \caption{Ratio functions (data/simulation) for hit charge vs.\ $\theta_{YZ}$. The solid lines are the values of the ratio in each bin, and the dashed lines are the spline fits. The bin at $1.4 < \theta_{YZ} < \pi/2~\text{rad}$ is extrapolated as described in the text.}
    \label{fig:prop_vs_thetaYZ_ratio}
\end{figure}

Figure~\ref{fig:prop_vs_thetaXZ_ratio} shows the ratio of data to simulation for the corrected hit charges and widths as a function of $\theta_{XZ}$, including the extrapolation to the high-angle region.
Figure~\ref{fig:prop_vs_thetaYZ_ratio} shows the ratio for the corrected hit charges as a function of $\theta_{YZ}$.
The hit width is not expected to vary as a function of $\theta_{YZ}$, and we find that this is true in our data to within 2\% (measured in the angular range up to $\theta_{YZ}$ of 1.3, after which saturation effects lead to non-gaussian waveforms which lead to biased width estimates).
For this reason we do not extract the ratio of the hit widths as a function of $\theta_{YZ}$ and do not apply this as part of our detector systematic uncertainties.

\FloatBarrier

\section{Wire Waveform Modification}
\label{sec:wiremod}

The functions based on the measured data to simulation ratios extracted in Section~\ref{sec:ratios} are used to modify the wire waveforms in simulated neutrino interaction events, effectively varying the detector response.
First, the wire signal regions are divided into Gaussian sub-regions in the drift time dimension that can be modified independently, again using the reconstructed hits.
This division is important because a single waveform can include overlapping charge from multiple particles with different kinematics that should be modified in different ways.
Additionally, because the simulated signals are overlaid on unbiased cosmic data as described in Section~\ref{sec:samples}, the algorithim must distinguish the simulation-dominated portions of the waveforms from the data-dominated portions.

Each wire signal region can be described as the sum of one or more Gaussian functions each with three parameters: peak position in time ticks, an integrated charge, and a width in time.
For each simulated energy deposit in the event, the projected position of the corresponding signal on each wire plane is computed after accounting for local nonuniformities in the electric field.
In this way simulated energy deposits are associated with the Gaussian regions that match their projected position.
The scale factors that are applied to the wire waveforms are based on the truth information of the simulated energy deposits matched to that portion of the waveform.
The individual simulated energy deposits each have an associated amount of energy as well as a start and an end position.
The $x$, $y$, and $z$ positions of the energy deposit are calculated as the average of the start and end positions; the angular variables $\theta_{XZ}$ and $\theta_{YZ}$ are computed using the definition in Equation~\ref{eqn:def_thetas}. 
The simulation modification functions derived in Section~\ref{sec:ratios} are used to obtain a charge and width scale factor for each energy deposit.
The hit charge and width scale factors for each Gaussian region of the wires are computed as the energy-weighted average of the scale factors over the associated set of energy deposits. For example, the scale factor $R$ for hit widths as a function of $x$ is given by
\begin{equation}
    R = \frac{\sum_i E_i \cdot R_\sigma(x_i)}{\sum_i E_i}
\end{equation}
    where the sums are over the set of energy deposits contributing to the Gaussian region, $E_i$ is the energy of the $i^\text{th}$ energy deposit, and $R_\sigma(x_i)$ is the spline fit for the hit widths as a function of $x$ from Figure~\ref{fig:prop_vs_x_ratio} evaluated at the $x$ position of the $i^\text{th}$ energy deposit.
The scale factors are set to unity if the Gaussian region has total charge greater than 80 units but less than 0.3~MeV of deposited energy associated with it.
This prevents small amounts of simulated charge from modifying cosmic-dominated regions of the waveforms.

\begin{figure}[ht]
    \centering
    \includegraphics[width=0.95\columnwidth]{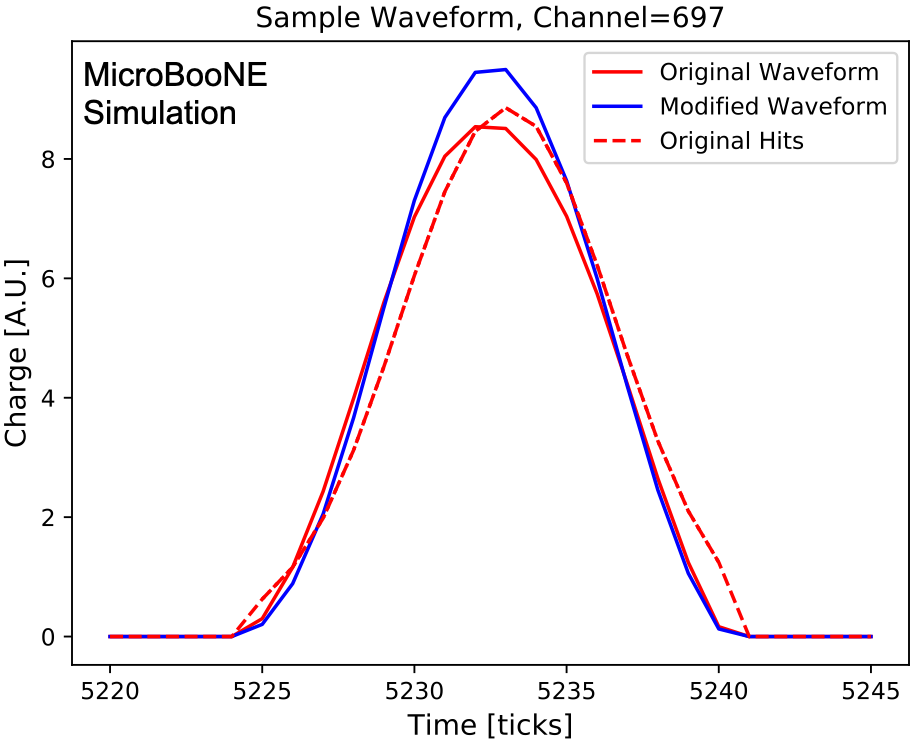}
    \includegraphics[width=0.95\columnwidth]{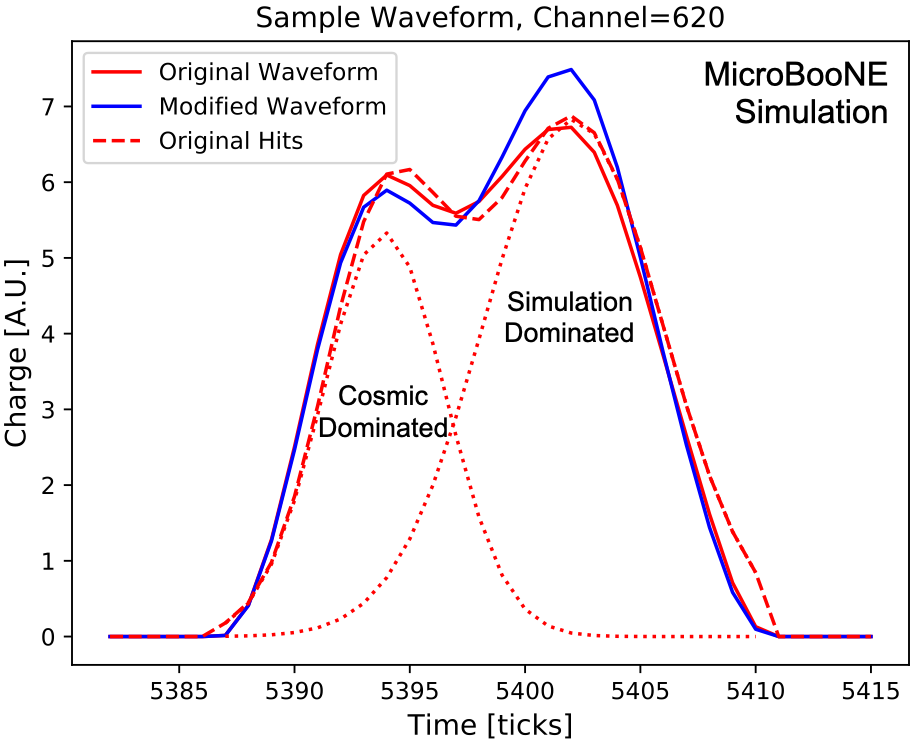}
    \caption{Examples of modified waveforms. The top graph shows a simple example where the wire signal region is well-described by a single Gaussian function. The bottom graph shows a case where one portion of the waveform is associated to simulated charge while the other is associated with cosmic data charge. Here, the simulation-dominated portion of the waveform is modified but the cosmic-dominated portion is not.}
    \label{fig:wiremod_examples}
\end{figure}

Finally, the above information is used to modify the overall waveform to have the desired integrated charge and width. This is accomplished by modifying the waveform at each time tick using the following procedure.
The original waveform is approximated by adding together the Gaussian functions that describe each region with their original parameters (mean time tick $t_0$, width $\sigma$, and integrated charge $Q$).
Similarly, the desired post-modification waveform is approximated by adding together the Gaussian functions with the same mean time tick but with modified charge $Q'$ and width $\sigma'$ based on their computed scale factors.
At each tick, the waveform is scaled by
\begin{equation}
    \text{scale}(t) = \frac{\sum_j \text{Gaus}(t; t_j, Q'_j, \sigma'_j)}{\sum_j \text{Gaus}(t; t_j, Q_j, \sigma_j)}
\end{equation}
where
\begin{equation}
\text{Gaus}(t; t_0, Q, \sigma) = \frac{Q}{\sqrt{2\pi} \, \sigma} \exp \left( - \frac{(t-t_0)^2}{2 \sigma^2} \right)
\end{equation}
with sums over the Gaussian region(s) within the relevant wire signal region.
Figure~\ref{fig:wiremod_examples} shows two examples of how this procedure modifies the waveforms.
The final result of running this procedure over an event is a new set of wire waveforms, where signals from simulated charge have been modified but signals from the cosmic data overlay are unchanged.
Waveform modifications are performed separately in each of the geometric variables, all in the manner described above for $x$. This results in one set of modified events for each of $x$, $(y,z)$, $\theta_{XZ}$, and $\theta_{YZ}$.

In order to validate this method, a closure test was performed using a simulation event sample in which the waveforms were modified in accordance with the ratios extracted above, and in which the hit properties were then re-measured.
The hit properties in the modified simulation are predicted exactly using the ratios the modification was based on, and the results show agreement within $\pm 2\%$ of those expectations in all variables.

An additional validation test was performed to demonstrate that this method can reproduce the behavior of a variation in a known detector model parameter. The variation used in this test case was a 50\% decrease in the longitudinal diffusion constant, consistent with the difference between the value measured in MicroBooNE data compared to the value used in the MicroBooNE simulation~\cite{diffusion}.
A sample of simulated ACPT muons was produced with this decreased diffusion constant, and used in place of detector data to extract a set of ratio functions that encapsulate the difference between the diffusion variation and the nominal simulation.
These ratio functions were then used to modify waveforms, according to the procedure described above, in a sample of simulated neutrino interactions (with nominal diffusion).
Finally, this wire-modified sample was compared to a sample of neutrinos generated with the diffusion constant modified in the initial simulation. We found the wire-modified sample to faithfully reproduce features of the diffusion-modified simulation across a range of low- and high-level reconstructed variables.

\FloatBarrier

\section{Uncertainties on Physics Observables}
\label{sec:impacts}

Post-modification simulated event samples for each of the variables $x$, $(y, z)$, $\theta_{XZ}$, and $\theta_{YZ}$ agree better with the data from the MicroBooNE detector in specific ways related to the wire response as a function of that variable.
This section details how small-statistics samples of simulated events with modified waveforms can be used to quantify any bias due to the detector mis-modeling in the nominal simulation, and how that bias can be included in the quoted systematic uncertainties.
The principle is that the difference between the nominal simulation and the modified simulations for each variable is used as the estimate of the corresponding bias. For most current MicroBooNE analyses, the bias is not corrected and is instead used as the estimated systematic uncertainty.

The wire modifications are determined based only on a sample of cosmic muons. As an example of general applicability, this section discusses applying them for evaluating systematic uncertainties on electromagnetic showers, objects very different from the charged particle tracks from which the wire modifications were derived.

\begin{figure}[h]
    \centering
    \includegraphics[width=\columnwidth]{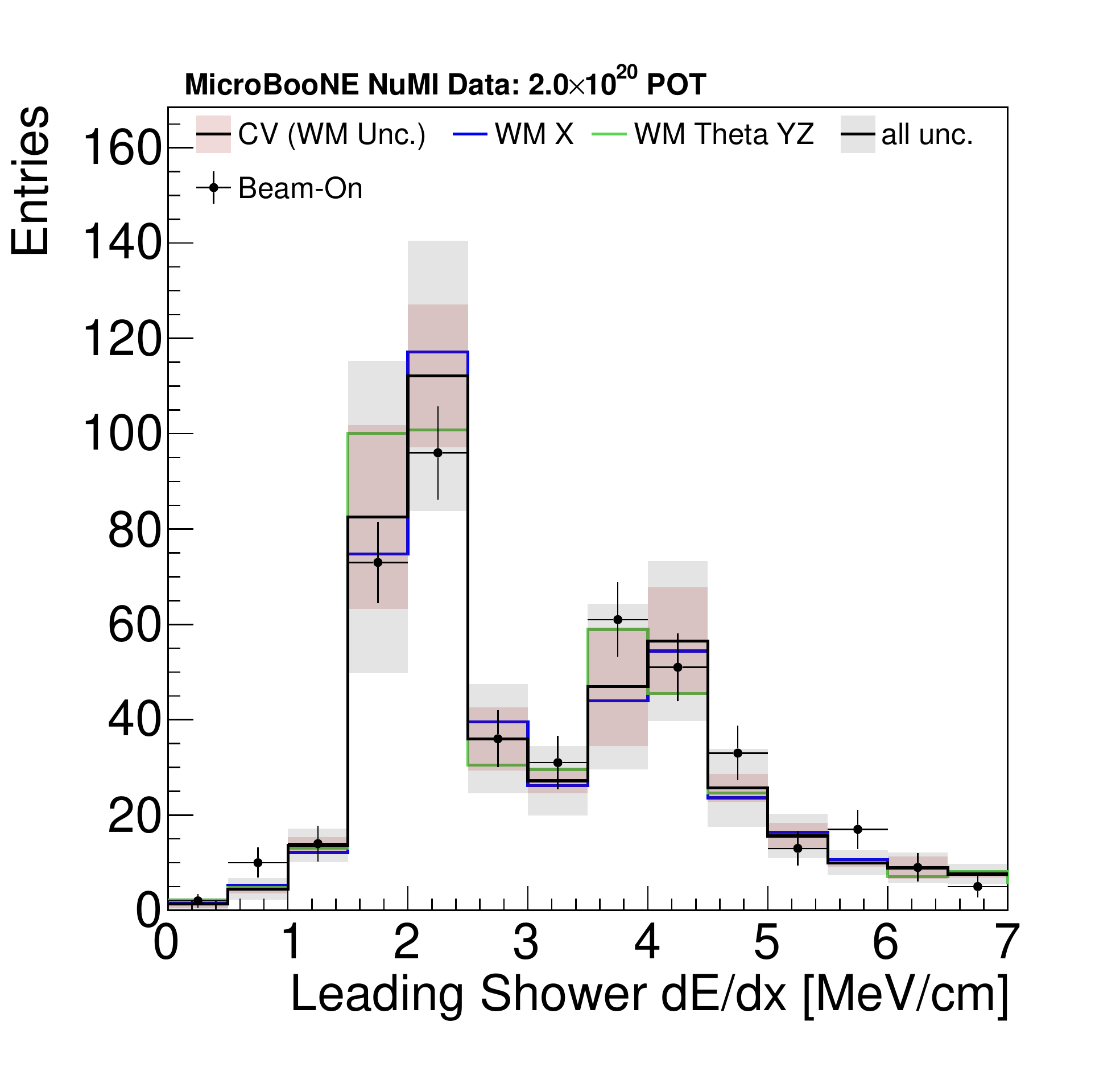}
    \caption{Distribution of the shower $dE/dx$ using NuMI beam data (points) and central value (CV) simulation (black line). The distribution for the simulation modified based on the detector response as a function of $x$ (blue line) and $\theta_{YZ}$ (green line) are also shown. The red band indicates the uncertainty from the wire modification alone (with all wire modification uncertainties included). The gray band indicates the full uncertainty, including other detector uncertainties as well as uncertainties on the neutrino flux and the interaction model. The bands represent the uncertainty on the number of events in that bin, calculated using Equation~\ref{eq:unc}, and are symmetric. The error bars on the data are statistical only.}
    \label{fig:nuMidEdX}
\end{figure}

For this study, two event samples are considered.
The first is a sample of single-shower events which are electron neutrino candidates from NuMI beam data~\cite{krishan}. 
For these showers, the energy loss per unit length, $dE/dx$, in the initial segment of the shower is measured. Electrons at the relevant energy scale will deposit energy as a minimum ionizing particle (2.1~MeV/cm), whereas photons produce showers primarily by pair production which will deposit twice as much energy per unit length (4.2~MeV/cm).
The measured $dE/dx$ of the trunks of these showers is shown in Figure~\ref{fig:nuMidEdX}, with the expected two contributions from electrons and photons.

\begin{figure}[h]
    \centering
    \includegraphics[width=\columnwidth]{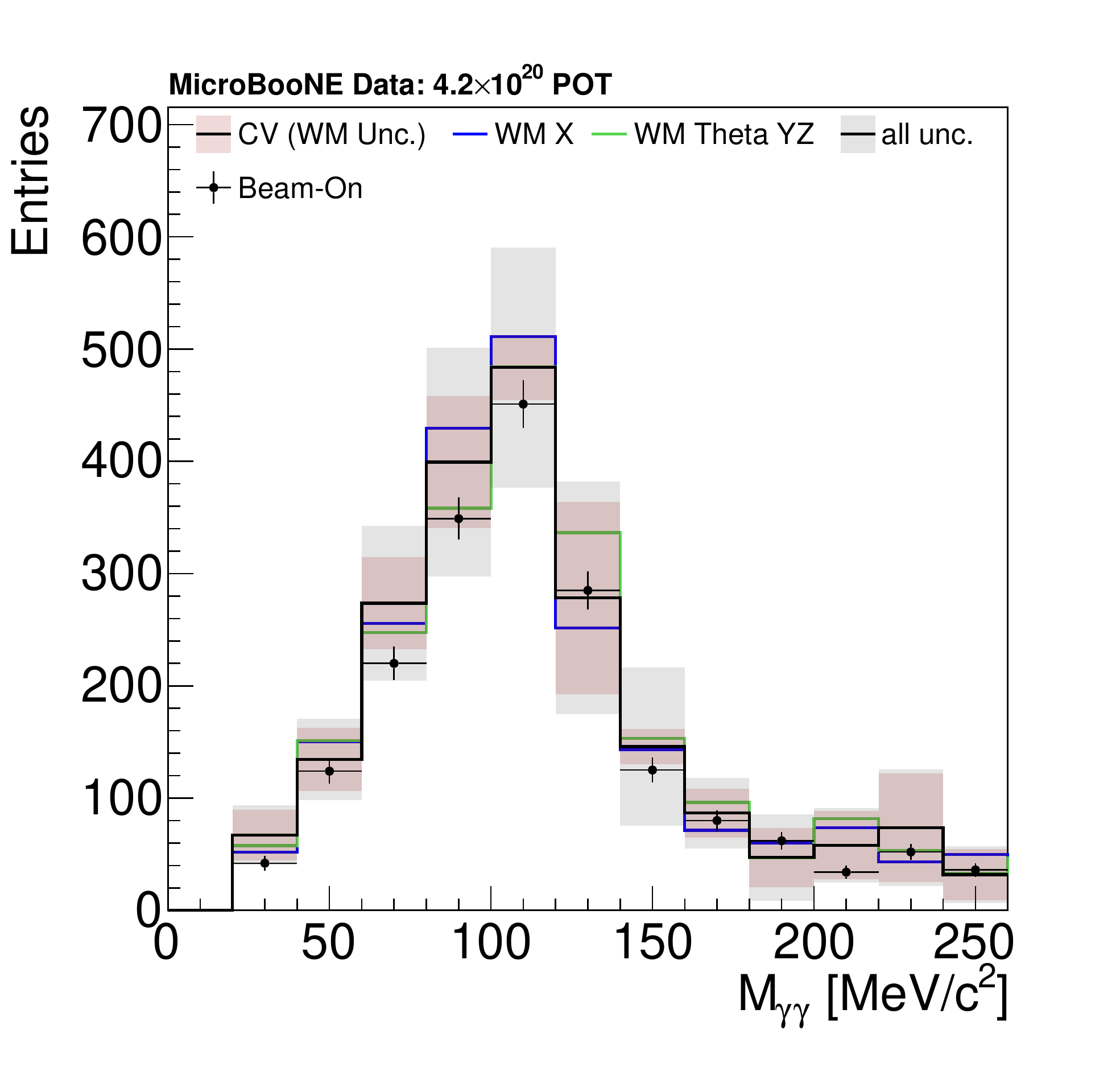}
    \caption{Measured diphoton invariant mass distribution using BNB beam data (points) and central value (CV) simulation (black line), prior to additional shower energy corrections. The distribution for the simulation modified based on the detector response as a function of $x$ (blue line) and $\theta_{YZ}$ (green line) are also shown. The red band indicates the uncertainty from the wire modification alone (with all wire modification uncertainties included).  The gray band indicates the full uncertainty, including other detector uncertainties, as well as uncertainties on the neutrino flux and the interaction model. The bands represent the uncertainty on the number of events in that bin, calculated using Equation~\ref{eq:unc}, and are symmetric. The error bars on the data are statistical only.}
    \label{fig:bnbPi0}
\end{figure}

The second sample is of events with two reconstructed photons, for which the primary production mechanism in MicroBooNE is neutral pion decay. This sample uses data from the BNB beam.
For each event in this sample, the diphoton invariant mass is calculated, as shown in Figure~\ref{fig:bnbPi0}. The shower energies are not corrected for known energy losses, such as shower clustering inefficiencies, so the invariant mass does not directly measure the neutral pion mass. However, this effect is present in both data and simulation.

The overall distributions of the $e/\gamma$ $dE/dx$ and diphoton invariant mass observables are subject to uncertainties from a range of sources. These include uncertainties in the flux and neutrino interaction model, but these uncertainties primarily manifest as normalization changes in the total number of events, or, in the case of the $e/\gamma$ $dE/dx$, relative normalization differences in the low (electron) and high (photon) ionization peaks.
The reconstructed positions and widths of the $dE/dx$ and $M_{\gamma\gamma}$ peaks are primarily driven by the detector response model, which is calibrated via the absolute charge scale measurement~\cite{calib}.
Errors in the response model lead to shifts in these distributions. Changes to the amplitudes and widths of the waveforms will change the measured amount of charge---even leading to charge falling below hit reconstruction thresholds---and so change the measurement of $dE/dx$, or lead to non-linear losses or gains in shower energy reconstruction.
Therefore, this study specifically looks at the peak positions and widths in order to evaluate the impact of the wire waveform modification procedure on these two distributions.

The mean and width, as measured using the RMS, of each of the peaks are calculated from unbinned data and simulation. The range that is used for each is given in Table~\ref{tab:peak_ranges}.
The systematic uncertainty on the simulation is calculated over the variations $s$ as
\begin{equation}
    \sigma_p = \sqrt{\sum_s(p_s - p_\text{CV})^2}
    \label{eq:unc}
\end{equation}
where $p_s$ and $p_\text{CV}$ are the parameters (either mean or RMS) estimated from each modified sample and the central value simulation, respectively.
The statistical uncertainty on the data is estimated assuming a Gaussian distribution.
The best-fit peak means and widths in the data and simulation and their uncertainties are summarized in Table~\ref{tab:meansAndWidths}. The wire modifications induce changes in the peak means and widths in simulation typically in the range of 1--2\%, though as large as $6\%$ in the case of the diphoton invariant mass width. These variations are consistent with the magnitude of the observed differences between the data and the simulation, suggesting that systematic uncertainties derived from this method are reasonable and not significantly overestimated.

\begin{table}[ht]
    \centering
    \begin{tabular}{|c|c|}
        \hline
        Value & Peak Range \\
        \hline
        e$^-$ $dE/dx$ & 1.75--3.0 MeV/cm \\
        $\gamma$ $dE/dx$ & 3.5--5 MeV/cm \\
        $M_{\gamma\gamma}$ & 20--200 MeV/c$^2$ \\
        \hline
    \end{tabular}
    \caption{Table summarizing the ranges used in calculating the means and widths of the peaks in the $dE/dx$ and diphoton invariant mass distributions.}
    \label{tab:peak_ranges}
\end{table}

\begin{table}[ht]
    \centering
    \begin{tabular}{ |c|c|c| }
        \hline
        Value & Data & MC \\
        \hline
        e$^-$ $dE/dx$ mean [MeV/cm] & $2.17 \pm 0.02$ & $2.15 \pm 0.05$ \\
        e$^-$ $dE/dx$ width [MeV/cm] & $0.342 \pm 0.017$  & $0.326 \pm 0.005$ \\
        $\gamma$ $dE/dx$ mean [MeV/cm] & $4.10 \pm 0.03$ & $4.08 \pm 0.05$ \\
        $\gamma$ $dE/dx$ width [MeV/cm] & $0.425 \pm 0.024$ & $0.423 \pm 0.010$\\
        $M_{\gamma\gamma}$ mean [MeV/c$^2$] & $106.5 \pm 0.9$ & $105.8 \pm 2.3$ \\
        $M_{\gamma\gamma}$ width [MeV/c$^2$] & $35.4 \pm 0.6$ & $36.6 \pm 2.3$ \\
        \hline
    \end{tabular}
    \caption{Table summarizing the mean and width of each of the peaks in the $dE/dx$ and diphoton invariant mass distributions. The data uncertainties are statistical, and the MC uncertainties are derived from the wire modified samples.}
    \label{tab:meansAndWidths}
\end{table}

\section{Future Work}
\label{sec:future}

The methods described in this paper have been used to estimate the impact of detector response uncertainties in MicroBooNE physics analyses.
There are a number of potential improvements and extensions possible.
The method could be expanded to describe the dependence on local ionization density.
This would require a sample of particles with varying energy deposition profiles, such as protons, with well-understood kinematic distributions that are similar between data and simulation.
Additionally, the dependence of the hit properties on the variables shown in this paper were shown to be separable from each other, except for the $y$ and $z$ position dependence which have strong correlations. The remaining correlations are known to be small, but in principle the dependencies could be measured simultaneously across more than two variables. Considering correlations in this way could further reduce the uncertainties on physics observables.
Finally, rather than taking the full difference between data and simulation, the methods described here could be used to correct the nominal simulation with the residual uncertainty on that correction being taken as the uncertainty.  This was not deemed necessary for recent MicroBooNE physics analyses, but could be employed if detector uncertainties became dominant.

\section{Summary and Conclusions}
\label{sec:summary}

This paper presents a novel method for applying data-driven modifications to simulated LArTPC wire waveforms.
The technique is based on comparisons between the properties of Gaussian hits fitted to the wire waveforms in data and simulation as  functions of the relevant variables: $x$, $(y,z)$, $\theta_{XZ}$, and $\theta_{YZ}$. The differences in waveform properties between data and simulation are used to modify simulated events, which are then used to quantify systematic differences in reconstructed variables.
This method is agnostic to the details of the simulation detector model and can capture mismodelling in known effects as well as unknown contributions not included in any model.
Compared to generating modified event samples repeating the full simulation with modified detector physics models, this method is more robust against underlying model assumptions and significantly more computationally efficient.

This paper has also shown how uncertainties on physics observables can be evaluated with this method using two MicroBooNE analyses as examples.
From this study, it was found that the wire waveform modification method leads to variations in electromagnetic shower-based observables consistent with the small differences between data and simulation, despite having been developed exclusively using cosmic muon tracks.
The method described here is generally applicable to wire-based noble liquid TPC detectors assuming the presence of a well-understood source for calibration samples with sufficient statistics. Such will be the case in the detectors of the Short Baseline Neutrino program at Fermilab, which, like MicroBooNE should have plentiful samples of cosmic-ray muon tracks, and in the case of the Deep Underground Neutrino Experiment far detector where laser or radioactive source calibration samples could be used to perform similar studies. Similar methods may be used for LArTPCs with different signal formation or readout mechanisms, though the applicability to these detector readout designs would have to be studied.

\begin{acknowledgements}
This document was prepared by the MicroBooNE collaboration using the resources of the Fermi National Accelerator Laboratory (Fermilab), a U.S.\ Department of Energy, Office of Science, HEP User Facility. Fermilab is managed by Fermi Research Alliance, LLC (FRA), acting under Contract No.\ DE-AC02-07CH11359.  MicroBooNE is supported by the following: the U.S.\ Department of Energy, Office of Science, Offices of High Energy Physics and Nuclear Physics; the U.S. National Science Foundation; the Swiss National Science Foundation; the Science and Technology Facilities Council (STFC), part of the United Kingdom Research and Innovation; the Royal Society (United Kingdom); and The European Union’s Horizon 2020 Marie Sklodowska-Curie Actions. Additional support for the laser calibration system and cosmic ray tagger was provided by the Albert Einstein Center for Fundamental Physics, Bern, Switzerland. We also acknowledge the contributions of technical and scientific staff to the design, construction, and operation of the MicroBooNE detector as well as the contributions of past collaborators to the development of MicroBooNE analyses, without whom this work would not have been possible.
\end{acknowledgements}

\end{document}